\documentclass[pra,twocolumn,groupedaddress]{revtex4-1}
\usepackage[english]{babel}
\usepackage{graphicx}
\usepackage{dcolumn}
\usepackage{bm}
\usepackage{epsfig}
\usepackage{amsmath}
\usepackage{amsfonts}
\usepackage{amssymb}
\usepackage{color}
\usepackage{bbm}
\usepackage{braket}
\usepackage{float}
\usepackage{enumitem}
\usepackage{tensor}
\usepackage{xcolor}
\usepackage{soul}

\usepackage[colorlinks=true,citecolor=blue,linkcolor=blue,urlcolor=blue]{hyperref}

\def\mathbi#1{\textbf{\em #1}}

\begin{document}

\title{Hybrid interfaces at the single quantum level in fluorescent molecules}

\author{Daniele De Bernardis$^{1}$, Hugo Levy-Falk$^{1}$, Elena Fanella$^{1,3}$,  Rocco Duquennoy$^{1}$, Valerio Digiorgio$^{4}$,  Giacomo Scalari$^{4}$, Maja Colautti$^{1,2}$, and Costanza Toninelli$^{1,2}$}
\affiliation{$^1$National Institute of Optics (CNR-INO), c/o LENS via Nello Carrara 1, Sesto F.no 50019, Italy}
\affiliation{$^2$European Laboratory for Non-Linear Spectroscopy (LENS), Via Nello Carrara 1, Sesto F.no 50019, Italy}
\affiliation{$^3$Physics Department, University of Naples, Via Cinthia 21, Fuorigrotta 80126, Italy}
\affiliation{$^4$ Institute for Quantum Electronics, ETH Zürich, 8093 Zürich, Switzerland.}
\date{\today}

\begin{abstract} 
We theoretically investigate a single fluorescent molecule as a hybrid quantum optical device, in which multiple external laser sources exert control of the vibronic states.
In the high-saturation regime, a coherent interaction is established between the vibrational and electronic degrees of freedom, and molecules can simulate several cavity QED models, whereby a specific vibrational mode plays the role of the cavity mode.
Focusing on the specific example where the system is turned into an analogue simulator of the quantum Rabi model, the steady state exhibits vibrational bi-modality resulting in a statistical mixture of highly non-classical vibronic cat states.
Applying our paradigm to molecules with prominent spatial asymmetry and combining an optical excitation with a THz(IR) driving, the system can be turned into a single photon transducer.
Two possible implementations are discussed based on the coupling to a subwavelength THz patch antenna or a resonant metamaterial. 
In a nutshell, this work assesses the role of molecules as an optomechanical quantum toolbox for creating hybrid entangled states of electrons, photons, and vibrations, hence enabling frequency conversion over very different energy scales.
\end{abstract}
 
\maketitle

\section{Introduction}
\label{sec:intro}


Molecules are naturally complex systems with transitions at different energy scales, corresponding to the excitation of vibrational, spin, or electronic degrees of freedom. This articulated energy diagram can be controlled and synthetically engineered, making the molecular platform increasingly attractive also for the development of quantum technologies \cite{gaita-arino_molecular_2019, mitra_quantum_2022, toninelli_single_2021}. 

At low temperatures, the lifetime-limited optical transition of single polycyclic aromatic hydrocarbon (PAH) fluorescent molecules is exploited to generate indistinguishable photons, or for quantum sensing applications \cite{lombardi_triggered_2021, duquennoy_real-time_2022, moradi_matrixinduced_2019,esteso_quantum_2023}. Even at room temperature, the coherence of vibrational wavepackets has been observed, probing single molecules by means of ultrafast spectroscopy \cite{brinks_visualizing_2010, hildner_femtosecond_2011, anderson_two-color_2018, velez_preparation_2019}. 

At the intersection between these regimes, the field of molecular optomechanics aims to exploit the interaction between optical and mechanical degrees of freedom. 
A seminal experimental work in this direction is the realization of the single-molecule optical transistor in Ref. \cite{hwang_single-molecule_2009}, or the more recent theoretical results described in Ref. \cite{schmidt_quantum_2016, schmidt_linking_2017}, enabling the IR to optical transduction based on a doubly-resonant surface-enhanced-Raman-scattering effect observed in \cite{chen_continuous-wave_2021, xomalis_detecting_2021, esteban_molecular_2022, chikkaraddy_single-molecule_2023}.

In this work, we theoretically show that in the high-saturation regime of the optical transition, quantum coherence between vibrational and electronic states can be established, leading to the formation of highly non-classical vibrational-electronic (vibronic) entangled states.
The system becomes an analogue of a cavity QED system, where the electronic two-level transition plays the role of the two-level atom and the vibrational mode represents the cavity bosonic mode.
The type of coherent interaction between them can be engineered by controlled light beams at different frequencies and in different spectral regimes, simulating quantum-Rabi-like Hamiltonians, with a similar approach to what was discussed in Ref. \cite{poyatos_quantum_1996} (albeit in the context of trapped ions reservoir engineering).
We hence present the molecule's multi-frequency driving as a \emph{quantum toolbox} to tailor the dynamics and the steady state of the vibronic degrees of freedom.
In this way, we explicitly show how molecular vibrations can span interesting quantum states, such as Schr\"odinger cat states, potentially interesting for quantum computing \cite{ralph_quantum_2003, gilchrist_schrodinger_2004, gravina_critical_2023} and quantum metrology \cite{munro_weak-force_2002}.

The time scales set by the fast (picosecond) vibrational relaxation push the system's dynamics to essentially ultra-fast time scales, making it hard to have direct measurements in conventional quantum optical setups \cite{toninelli_single_2021}.
Natural observables such as hybrid vibronic vacuum Rabi oscillations \cite{haroche_exploring_2006} represent prohibitive measurements. However, the molecule's emitted fluorescence is sensibly altered in this regime, becoming indirect evidence of the coherently populated vibrational levels.
Even more striking, after establishing a finite vibronic population using two lasers at different frequencies, the addition of a supplemental bias laser can induce quantum jumps \cite{blatt_quantum_1988} between vibrational states, following slow dynamics.
This analysis, without being fully exhaustive, is sufficient to show that hybrid vibronic features can also be explored with conventional quantum optical setups. 

Finally, when a laser at optical frequencies is combined with a source in a lower-frequency regime, like at THz frequencies, this platform implements a quantum transducer.
The map over a cavity QED model provides now a simple language to identify and characterize the optimal conditions and figures of merit to reach transduction at the single-photon level.
Leveraging on the versatility of fluorescent molecules with respect to the coupling with complex photonic structures \cite{kuhn_enhancement_2006, toninelli_scanning_2010, wang_coherent_2017, luo_electrically_2019, schadler_electrical_2019, esteso_quantum_2023}, we propose two concrete experimental implementations based on currently available patch antenna quantum cascade lasers and resonant metamaterials that could potentially realize a single photon THz-to-optical transducer (and thus THz single photon detector).

Overall, our work establishes a single fluorescent molecule as the smallest and fastest optomechanical device available up to now, furthermore, the only one that can fully achieve the non-perturbative(/non-linear) single-photon regime \cite{rabl_photon_2011, nunnenkamp_single-photon_2011}.

The article is structured as follows.
In Sec. \ref{sec:model}, we present the general model describing the molecular setup.
In Sec. \ref{sec:rrqe}, we show how to engineer the interactions between the electronic and vibrational levels of the molecules, simulating various cavity QED toy models.
In Sec. \ref{sec:blinking}, we explore a bi-modal vibrational steady-state using the quantum trajectory method.
In Sec. \ref{sec:trans}, we develop the general theory of a molecular THz-to-optical photon transducer.
In Sec. \ref{sec:implementation_THZ}, we present the possible experimental implementation of the transducer with THz patch antennas or resonant metamaterials.
Finally, in Sec. \ref{sec:conclusion}, we summarize our conclusions.

\section{Model}
\label{sec:model}

\begin{figure}
    \centering
    \includegraphics[width=\columnwidth]{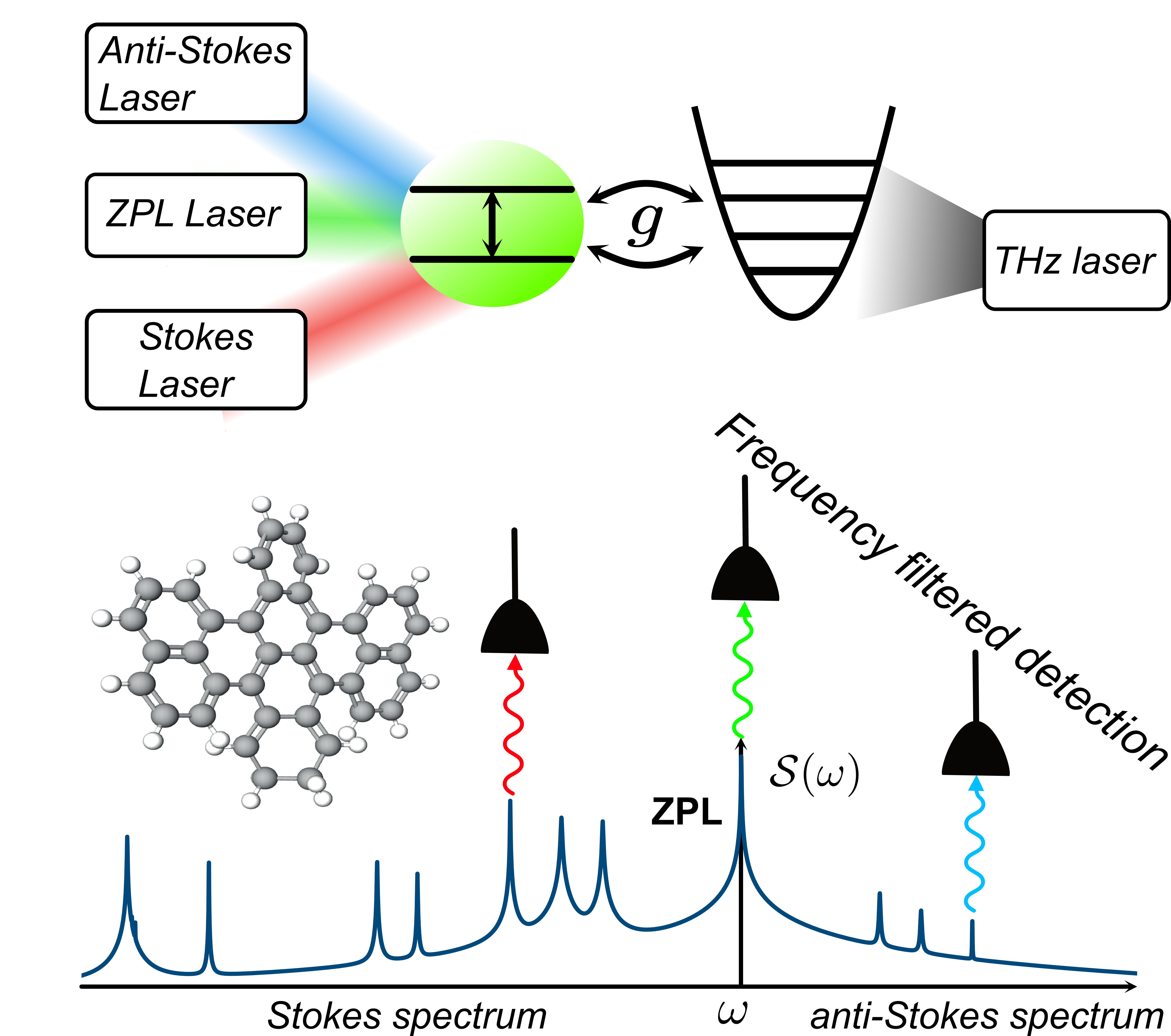}
    \caption{Schematic representation of the considered setup. Multiple lasers drive a fluorescent molecule at different frequencies in the optical and THz range. The emitted fluorescence is in the optical domain, and its spectrum consists of several Raman peaks (Stokes and anti-Stokes) displaced around the molecule's HOMO-LUMO transition frequency, dubbed as zero-phonon line. A frequency-filtered detection is performed to investigate the state of the molecule. The spectrum sketched in the top is a fit from Eq. \eqref{eq:multi_mode_S} using a set of parameters taken from Ref. \cite{zirkelbach_high-resolution_2022} (see Apps. \ref{app:frequency_filter}-\ref{app:generalization_multiple} for further details).
    The DBT molecule is shown here as a proxy of a single-photon emitter large organic molecule.}
    \label{fig:1}
\end{figure}

We consider a single molecule defined by an electronic transition between the molecular electronic ground $|g\rangle$ and excited states $|e\rangle$ (HOMO-LUMO transition), which frequency is $\omega_0 = \omega_e - \omega_g$ (here $\hbar\omega_g,\hbar\omega_e$ are the eigenenergies of the electronic ground and the excited state).
Due to the Born-Oppenheimer approximation, to each molecular orbital is associated a multi-dimensional potential energy surface (PES), describing the vibrational-mode dynamics \cite{born_zur_1927, di_bartolo_optical_2010, gurlek_small_2024}.
For a highly rigid molecule (see, for example, the dibenzoterrylene (DBT) molecule \cite{toninelli_single_2021}, whose structure is represented in Fig. \ref{fig:1}), each PES has a well-defined minimum that can be approximated by a parabolic potential, making each vibrational mode behave as a harmonic oscillator with a specific frequency $\omega_{\rm v}$.
The displacement in the minima of the vibrational coordinate between the HOMO and LUMO PES gives rise to an energy shift quantified by $\varepsilon_1$. 

Considering a single vibrational degree of freedom, this construction results in a two-level system coupled through a conditional displacement with a single harmonic oscillator. 
The system Hamiltonian is given by the Holstein model \cite{holstein_studies_1959, spano_excitons_2006,clear_phonon-induced_2020,reitz_molecule-photon_2020, sommer_molecular_2021, zhang_nonlinear_2024}
\begin{equation}
    H_{\rm mol} = \hbar \omega_0 \hat{\sigma}^{\dag} \hat{\sigma} + \hbar \omega_{\rm v} \hat{b}^{\dag} \hat{b} + \varepsilon_1 (\hat{b} + \hat{b}^{\dag})\hat{\sigma}^{\dag}\hat{\sigma}.
\end{equation}
Here $\hat{\sigma} = |g \rangle \langle e |$ is the electronic rising operator, and $\hat{b}$ is the annihilation operator of a vibrational excitation within the PES mentioned above and fulfilling the Bosonic commutation relation $[\hat{b}, \hat{b}^{\dag}] = 1$.
To fulfill the Born-Oppenheimer approximation, the frequencies must respect the order $\omega_0 \gg \omega_{\rm v}$.
Moreover, we have assumed that the two PESs have minima with the same curvature. The general case is treated in Ref. \cite{zhang_nonlinear_2024}, where the Holstein Hamiltonian is corrected by an additional non-linear term. 

As already anticipated in the introduction in Sec. \ref{sec:intro}, our discussion is developed with a special focus on PAH molecules \cite{toninelli_single_2021}, whose vibrational levels are in the THz range, and the purely electronic transition varies from optical to near infrared frequencies.
For simplicity, in the following, we will hence assume the vibrational mode and the electronic transition to be in the THz and in the optical domain, respectively.

The molecule can be driven simultaneously by multiple optical lasers and a THz source with frequencies $\lbrace{ \omega_{\rm L_{\ell}} \rbrace}|_{(\ell=1,2\ldots N_{\rm L})}$, $\omega_{\rm THz}$, respectively, described by the time-dependent Hamiltonian contributions
\begin{equation}\label{eq:opt_drive_ham_normal}
    H_{\rm opt} = \sum_{\ell=1}^{N_{\rm L}} \hbar \Omega_{\rm L_{\ell}}\cos \left(\omega_{\rm L_{\ell}} t \right) \left( e^{i\phi_{\ell}}\hat{\sigma} + e^{-i\phi_{\ell}}\hat{\sigma}^{\dag} \right),
\end{equation}
\begin{equation}\label{eq:thz_drive_ham_normal}
    H_{\rm THz} = \hbar \Omega_{\rm THz}\cos \left(\omega_{\rm THz} t \right) \left( e^{i\theta}\hat{b} + e^{-i\theta}\hat{b}^{\dag} \right).
\end{equation}
Here $\Omega_{\rm L_{\ell}}$, $\Omega_{\rm THz}$ are the $\ell$-th optical and THz Rabi frequencies and $\phi_{\ell}, \theta$ are arbitrary phases introduced for completeness. Without loss of generality, in this work, we fix $\phi_{\ell}=\theta=0$.

The setup presented above is pictorially represented in Fig. \ref{fig:1}.

\subsection{Polaron dressing: ZPL, Stokes and anti-Stokes side-bands}
\label{sec:polaron_dressing}
From nanomechanics to quantum dots and molecular cavity QED, a common technique to treat the coupling between optics and mechanics at the quantum optical level is the so-called \emph{polaron transformation} \cite{rabl_photon_2011,mccutcheon_general_2011,kirton_nonequilibrium_2013, holzinger_cooperative_2022, de_bernardis_relaxation_2023}.
Applying the unitary operator $U=\exp \left[ \eta (\hat{b} - \hat{b}^{\dag})\hat{\sigma}^{\dag}\hat{\sigma} \right]$ we transform the vibrational and electronic operators as follows \cite{holzinger_cooperative_2022}
\begin{align}\label{eq:polaron_mapping_operators}
    \hat{b} \mapsto \hat{b}_{\rm pol} =  \hat{b} - \eta \hat{\sigma}^{\dag}\hat{\sigma} && \hat{\sigma} \mapsto \hat{\sigma}_{\rm pol} = \hat{\mathcal{D}}(\eta)\hat{\sigma} ,
\end{align}
where $\eta = \varepsilon_1/(\hbar \omega_{\rm v})$ is the square root of the Franck-Condon factor \cite{clear_phonon-induced_2020} (see end of App. \ref{app:frequency_filter} for more details on this definition), and $\hat{\mathcal{D}}(\eta)$ is the displacement operator \cite{walls_quantum_2008}, that can be written as \cite{cahill_ordered_1969}
\begin{equation}\label{eq:displacement_normal_order_expansion}
    \hat{\mathcal{D}}(\eta) = e^{-\frac{\eta^2}{2}} \sum_{m, m' = 0}^{+\infty} \frac{(\eta \hat{b}^{\dag})^{m}}{m!}\frac{(-\eta \hat{b})^{m'}}{m'!}.
\end{equation}

The polaron frame is particularly useful since it diagonalizes the Holstein Hamiltonian \cite{holzinger_cooperative_2022}, disentangling the electronic and the vibration degrees of freedom
\begin{equation}\label{eq:Ham_mol_polaron}
    UH_{\rm mol}U^{\dag} = \hbar \omega_0 \hat{\sigma}^{\dag} \hat{\sigma} + \hbar \omega_{\rm v} \hat{b}^{\dag} \hat{b}.
\end{equation}
From here on, $\hat{\sigma}$ and $\hat{b}$ represent operators acting in the polaron basis.
All the information regarding the vibrational coupling is then moved to the driving terms, resulting in an optical Hamiltonian that is essentially equivalent to those of trapped ions \cite{leibfried_quantum_2003}
\begin{equation}\label{eq:ham_optical_trapped_ion}
    H_{\rm opt} \approx \sum_{\ell=1}^{N_{\rm L}} \frac{\hbar \Omega_{\rm L_{\ell}}}{2} \left( \hat{\sigma}e^{ -\eta (\hat{b} - \hat{b}^{\dag}) - i\omega_{\rm L_{\ell}} t} + {\rm h.c}  \right).
\end{equation}
For the moment, we have assumed the rotating-wave approximation (RWA) without too many justifications, but it will be clarified below.

The Hamiltonian in Eq. \eqref{eq:ham_optical_trapped_ion} can be interpreted as follows: Whenever an optical photon is absorbed or emitted, it induces a recoil in the molecular vibrations, hence displacing its vibration coordinate.
The amount of displacement in the recoil process is quantified by $\eta$, which takes the role of the Lamb-Dicke parameter \cite{cirac_laser_1992, leibfried_quantum_2003}, whose square is the Franck-Condon factor.
Depending on the specific vibrational mode, typical values for DBT in anthracene matrices are $\eta \sim 0.1 - 0.3$ \cite{clear_phonon-induced_2020, zirkelbach_high-resolution_2022}. 

The system is then in the so-called Lamb-Dicke regime, where $\eta < 1$, and the polaron interaction can be linearized at first order in $\eta$, discarding higher order contributions in the expansion of Eq. \eqref{eq:displacement_normal_order_expansion}. In Sec. \ref{sec:blinking}, we will see how higher order corrections can nevertheless play a role.
The polaron rising operator $\hat{\sigma}_{\rm pol}$ in Eq. \eqref{eq:polaron_mapping_operators}, in the interaction picture with respect to the free polaron Hamiltonian of Eq. \eqref{eq:Ham_mol_polaron}, can be reduced to only three contributions
\begin{equation}\label{eq:sigma_pol_frequency_decomp}
\begin{split}
    &\hat{\sigma}_{\rm pol}(t) \approx \hat{\sigma} e^{-i\omega_0 t} 
    \\
    & + \eta \left( \hat{\sigma}\hat{b}^{\dag}  e^{-i(\omega_0-\omega_{\rm v}) t} -\hat{\sigma}\hat{b} e^{-i(\omega_0+\omega_{\rm v}) t} \right) + O(\eta^2).
\end{split}
\end{equation}
These three terms represent the rising operator of the ZPL, the Stokes and anti-Stokes transition, respectively.
For this reason, from here on, we focus mostly on the case where we have at most three optical driving terms, $N_{\rm L}=3$. 
Their resonance and Rabi frequencies will be called ZPL, Stokes and anti-Stokes $\omega_{\rm L_{\ell}} = (\omega_{\rm zpl}, \omega_{\rm S}, \omega_{\rm AS})$, $\Omega_{\rm L_{\ell}} = ( \Omega_{\rm zpl},  \Omega_{\rm S},  \Omega_{\rm AS})$, with $\omega_{\rm zpl}\approx \omega_0,\, \omega_{\rm S}\approx \omega_0 - \omega_{\rm v},\, \omega_{\rm AS}\approx \omega_0 + \omega_{\rm v}$.

Given this structure, the RWA naturally applies to each ZPL, Stokes and anti-Stokes term by safely assuming $\Omega_{\rm L_{\ell}} \ll \omega_0, \omega_0\pm \omega_{\rm v}$, yielding $H_{\rm opt} \approx H_{\rm zpl} + H_{\rm S} + H_{\rm AS}$, where
\begin{equation}\label{eq:ham_drive_zpl}
    H_{\rm zpl} = \frac{\hbar \Omega_{\rm zpl}}{2}  \left( \hat{\sigma} e^{i\omega_{\rm zpl} t} + \hat{\sigma}^{\dag} e^{-i\omega_{\rm zpl} t} \right),
\end{equation}
\begin{equation}\label{eq:ham_drive_stokes}
    H_{\rm S} = \frac{\hbar g_{\rm S}}{2} \left( \hat{\sigma}\, \hat{b}^{\dag} e^{i\omega_{\rm S} t} + \hat{\sigma}^{\dag}\, \hat{b} e^{-i\omega_{\rm S} t} \right),
\end{equation}
\begin{equation}\label{eq:ham_drive_antistokes}
    H_{\rm AS} =  \frac{\hbar g_{\rm AS}}{2} \left( \hat{\sigma}\, \hat{b} e^{i\omega_{\rm AS} t} + \hat{\sigma}^{\dag}\, \hat{b}^{\dag} e^{-i\omega_{\rm AS} t} \right).
\end{equation}
Here, we have introduced the effective Stokes/anti-Stokes coupling constants
\begin{equation}\label{eq:g_S_definition}
    g_{\rm S,AS} = \eta \, \Omega_{\rm S,AS}.
\end{equation}
If the frequencies $\omega_0, \omega_{\rm S}, \omega_{\rm AS}$ are assumed to be well separated from each other, one can safely sum all these contributions to obtain the compact approximated form of Eq. \eqref{eq:ham_optical_trapped_ion}.
With the same level of approximation, we also obtain the THz drive, which reads 
\begin{equation}
    H_{\rm THz} \approx \frac{\hbar\Omega_{\rm THz}}{2}\left( \hat{b} e^{i\omega_{\rm THz} t} + \hat{b}^{\dag}e^{-i\omega_{\rm THz} t} \right).
\end{equation}

Recasting the system in these terms allows us to single out the ZPL, Stokes, and anti-Stokes side-bands and the resonant processes activating them. 
The Hamiltonian with optical driving presented above leads to a Raman side-band toolbox for the vibrational state manipulation in very close analogy to what was developed for trapped ions \cite{cirac_laser_1992,poyatos_quantum_1996,leibfried_quantum_2003} and is schematically represented in Fig. \ref{fig:1}, upper panel.
It is worth noticing that the same treatment can be applied as well to the optomechanical coupling between the molecule's electronic transition and the matrix's phonons \cite{clear_phonon-induced_2020, gurlek_engineering_2021}, leading to the same formal structure.

\subsection{Frequency filtering and detection}
\label{sec:filters}
In a typical quantum optics experiment with fluorescent molecules, the photon detection is performed by exploiting the rich multi-frequency emission spectrum.
Filtering away the light scattered at the driving frequencies and detecting fluorescence in a different spectral range allows for background-free detection and high signal-to-noise ratios, reaching the single-molecule regime \cite{orrit_single_1990}.
In Fig. \ref{fig:1} lower panel, we schematically show this concept, where the molecule's spectrum exhibits several vibrational modes Stokes side peaks and their respective anti-Stokes counterparts, while the detection focuses only on the ZPL and the (anti)Stokes of a specific mode.

Following the calculations outlined in App. \ref{app:frequency_filter}, the fluorescence (photon counting) rate detected through the filter is given by the formula:
\begin{equation}
    \Gamma_{\mathcal{F}} = p_{\rm click}\gamma_{0} \langle{ \hat{\mathrm{N}}_{\mathcal{F}} \rangle},
\end{equation}
where $p_{\rm click}$ is the probability of generating a photocurrent upon incidence of a single photon on the detector, and $\gamma_{0}$ is the ZPL spontaneous emission radiative rate.
For a narrow filter centered either on the ZPL, Stokes, or anti-Stokes side-bands, the filtered photon number operator $\hat{\mathrm{N}}_{\mathcal{F}}$ is identified by the corresponding side-band. In this way, we can define the ZPL, Stokes, and anti-Stokes photon number operator as
\begin{equation}\label{eq:N_photon_number_zpl_S_AS}
    \begin{split}
        \hat{\mathrm{N}}_{\rm zpl} &=  \hat{\sigma}^{\dag}\hat{\sigma},
        \\
     \hat{\mathrm{N}}_{\rm S}  & = \eta^2\hat{\sigma}^{\dag}\hat{\sigma} \left( 1 +  \hat{b}^{\dag}\hat{b}  \right),
    \\
     \hat{\mathrm{N}}_{\rm AS}  &= \eta^2  \hat{\sigma}^{\dag}\hat{\sigma} \, \hat{b}^{\dag}\hat{b}.
    \end{split}
\end{equation}
Here, we see that the Stokes and anti-Stokes photon numbers are multiplied by the Franck-Condon factor, $\eta^2$, resulting in a strong suppression (typically one or two order of magnitudes) of the detected Stokes/anti-Stokes shifted fluorescence with respect to the ZPL one.

The Stokes and anti-Stokes photon number operators are a combination of electronic and vibrational operators, bringing direct information about the molecular vibrational state through the emitted fluorescence.
We immediately see that the Stokes photon number is proportional to the ZPL occupation, $\langle{ \hat{\mathrm{N}}_{\rm S} \rangle} \sim \langle{ \hat{\sigma}^{\dag}\hat{\sigma} \rangle}$, indicating a finite fluorescence also when the vibrational mode is empty.
On contrary, the anti-Stokes photon number is zero $\langle{ \hat{\mathrm{N}}_{\rm AS} \rangle} = 0$ if the vibrational mode is in the vacuum state, with zero population $\langle{ \hat{b}^{\dag}\hat{b} \rangle} = 0$, becoming a clear unambiguous probe of the vibrational occupation.

\subsection{Dissipative dynamics}
\label{sec:master_eq}

A molecule trapped in a solid-state matrix interacts with the surrounding environment both electromagnetically and mechanically, so it has various dissipation channels where it can release its excitations.
The system is thus intrinsically driven-dissipative and described by a master equation for its density matrix $\hat{\rho}$ (see App. \ref{app:master} for further details):
\begin{equation}\label{eq:master_equation_general}
    \hbar \partial_t \hat{\rho} = \mathcal{L}_{H}(\hat{\rho} ) + \mathcal{L}_{\gamma_{0}}(\hat{\rho}) + \mathcal{L}_{\gamma_{\rm v}}(\hat{\rho}).
\end{equation}
The coherent dynamic (including the driving terms) is obtained by summing up all the contributions in the Hamiltonian as follows
\begin{equation}
    \mathcal{L}_{H}(\hat{\rho} ) = -i\left[ H_{\rm mol} + H_{\rm opt} + H_{\rm THz}, \hat{\rho}  \right],
\end{equation}
while the electronic and vibrational dissipation are described by the two Lindbladian terms
\begin{equation}\label{eq:Lindblad_opt_photon_loss}
    \mathcal{L}_{\gamma_{0}}(\hat{\rho}) \approx \frac{\hbar \gamma_{0}}{2}\left[ 2\hat{\sigma}\, \hat{\rho} \, \hat{\sigma}^{\dag} - \lbrace{ \hat{\sigma}^{\dag}\hat{\sigma},\, \hat{\rho} \rbrace} \right],
\end{equation}
\begin{equation}\label{eq:Lindblad_v}
    \mathcal{L}_{\gamma_{\rm v}}(\hat{\rho}) \approx \frac{\hbar \gamma_{\rm v}}{2}\left[ 2\hat{b}\, \hat{\rho} \, \hat{b}^{\dag} - \lbrace{ \hat{b}^{\dag}\hat{b}, \hat{\rho} \rbrace} \right].
\end{equation}
Here, the LUMO-HOMO electronic decay rate coincides with the spontaneous decay rate $\gamma_0$ while in principle the electronic decay is also affected by non-radiative losses. However, considering PAH molecules at cryogenic temperatures, the latter can be neglected \cite{bassler_generalized_2024}.
On the contrary, $\gamma_{\rm v}$ is the vibrational dissipation rate and is completely dominated by non-radiative processes, where the vibrational excitation decays into the phononic modes of the surrounding matrix.
For this reason, they typically exceed the electronic decay by a few order of magnitude $\gamma_{\rm v} \gg \gamma_{0}$ \cite{clear_phonon-induced_2020, toninelli_single_2021, zirkelbach_high-resolution_2022}, becoming the dominant frequency scale of the dissipative processes.
We take inspiration from Refs. \cite{clear_phonon-induced_2020, zirkelbach_high-resolution_2022} to fix the dissipation rates through the whole paper as
\begin{align}
    \gamma_{0}/(2\pi) \approx 40\,{\rm MHz} && \gamma_{\rm v}/(2\pi)\approx 10\, {\rm GHz},
\end{align}
holding at $T\approx 1\,$K \cite{clear_phonon-induced_2020}.
It is worth noticing that this hierarchy in the dissipations represents a major difference with the physics of trapped ions, where the vibrational-phononic modes have instead negligible losses with respect to the optical one \cite{poyatos_quantum_1996}.

The two expressions above in Eq. \eqref{eq:Lindblad_opt_photon_loss}-\eqref{eq:Lindblad_v} are approximated forms, as explained in detail in App. \ref{app:master}. 
In general, they should involve the polaron operator in Eq. \eqref{eq:polaron_mapping_operators}, accounting for the vibrational dressing of the electronic jump operator $\hat{\sigma}$. 
However, with the value of the electronic decay rate given above, $\gamma_{0}/(2\pi)\approx 40\,$MHz, all the contributions from the polaron expansion in Eq. \eqref{eq:displacement_normal_order_expansion}, scaling at least with $\eta^2 \ll 1$, are completely negligible \cite{holzinger_cooperative_2022,juan-delgado_addressing_2025}. 
Although the approximation for the vibrational dissipation is slightly different, as explained in App. \ref{app:master}, it yields the same conclusion.

\section{Hybrid quantum states engineering}
\label{sec:rrqe}
In this section, we analyze the dynamics of the system when continuously driven by different combinations of the Stokes, ZPL, and anti-Stokes lasers in Eqs. \eqref{eq:ham_drive_stokes}-\eqref{eq:ham_drive_zpl}-\eqref{eq:ham_drive_antistokes}, but always keeping the THz radiation at zero input, $\Omega_{\rm THz}=0$.

The main focus is to exploit this configuration to excite different types of vibrational steady states, following the dynamics of a vast class of generalized quantum Rabi models \cite{de_bernardis_tutorial_2024}.
Off-resonant pumping and stimulated resonant Raman scattering \cite{prince_stimulated_2017} are specific examples that fall into this framework, but the phenomenology can be arbitrarily extended by varying the parameter range and by playing with the possible combinations of laser drivings. 

In this way, we create a new toolbox for the coherent manipulation of vibronic degrees of freedom at the single quantum level, paving the way for multi-scale quantum engineering in molecules.

After composing a specific Hamiltonian, this is plugged into the open-dissipative dynamics given by the master equation in Eq. \eqref{eq:master_equation_general}, thus providing the complete description of the system's evolution.
Looking at the steady state of the system, $\partial_t \hat{\rho}_{ss} = 0$, we discuss some paradigmatic examples of this phenomenology.

\subsection{Stokes coupling}

\begin{figure}
    \centering
    \includegraphics[width=\columnwidth]{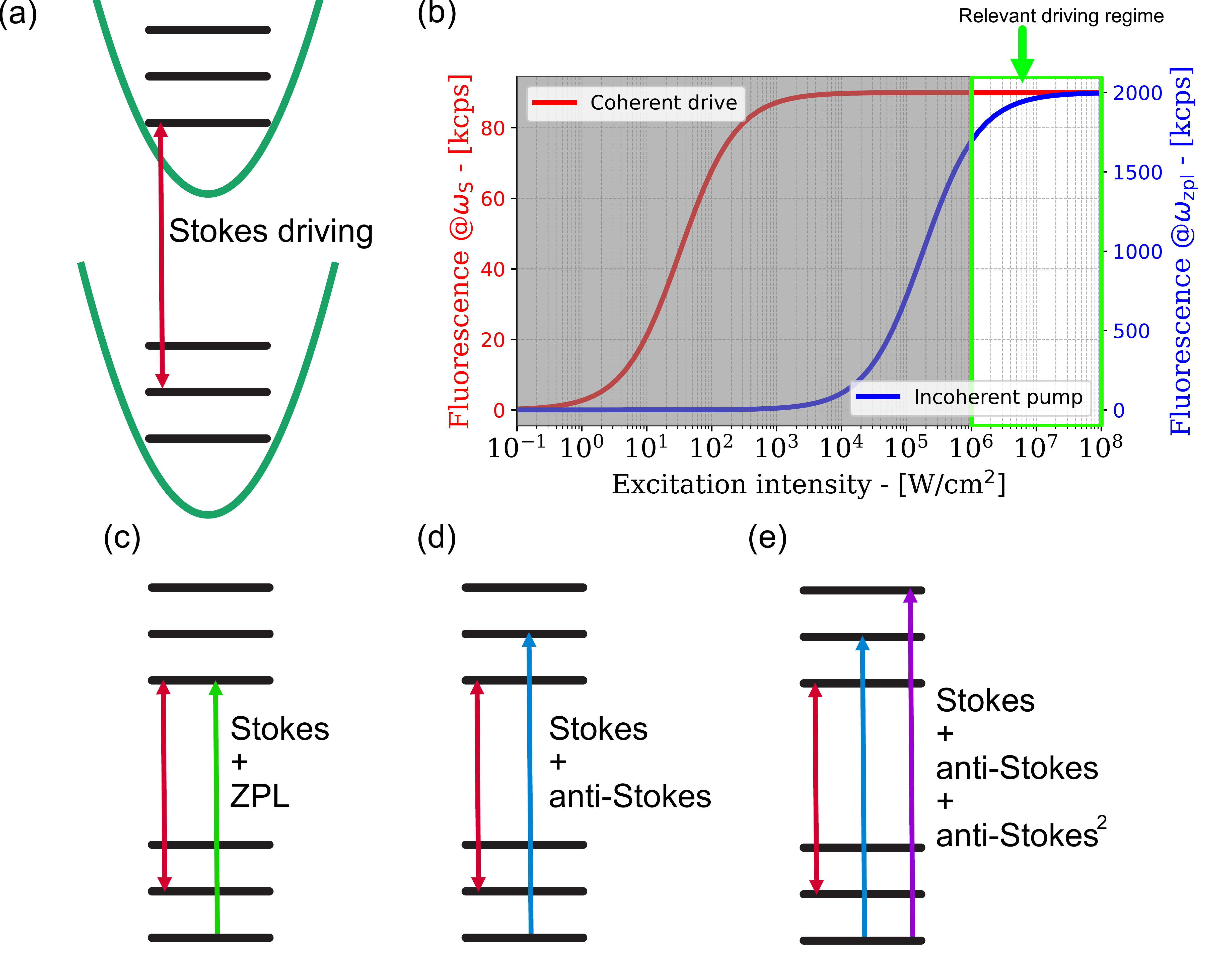}
    \caption{(a) Level scheme representation of the Stokes driving. (b) Photon counting rate of the fluorescence signal (see App. \ref{app:sat_curve}). Red solid line: ZPL resonant excitation and Stokes detection, as described by Eq. \eqref{eq:photo_rate_Stokes_zplexc}. Blue solid line: incoherent (off-resonant) excitation and ZPL detection, as described by Eq. \eqref{eq:photo_rate_incoh}. Parameters: $\gamma_{0}/(2\pi)=40\,$MHz, $\gamma_{\rm v}/(2\pi)=10\,$GHz, $\eta = 0.3$, $p_{\rm click}=0.05$. The electronic dipole transition strength is $\xi_{0} = 0.23\,$nm, resulting in $\Omega_{\rm zpl}^s/(2\pi)\approx 2.8\times 10^{-2}\,$GHz at saturation ($n_{\rm zpl}^s\approx 1$). The incoherent saturation anti-Stokes Rabi frequency is instead at $\Omega_{\rm AS}^s/(2\pi)\approx 2$GHz for the same parameters.
    Level scheme of combined (c) Stokes and ZPL drivings, (d) Stokes and anti-Stokes drivings, (e) Stokes, anti-Stokes and second order anti-Stokes (AS$^2$) drivings.
    }
    \label{fig:2}
\end{figure}

The fundamental building block for hybrid quantum states engineering is the Stokes driving in Eq. \eqref{eq:ham_drive_stokes}, which realizes a coherent conversion of the electronic excitation into a vibrational one (and vice-versa).
This basic level scheme is represented in Fig. \ref{fig:2}(a).
Switching to a rotating frame, in the polaron framework the molecular Hamiltonian takes the shape of the Jaynes-Cummings model \cite{larson_jaynescummings_2024}
\begin{equation}\label{eq:ham_JC}
\begin{split}
    H_{\rm mol} + H_{\rm S} &= H_{\rm JC}
    \\
    & = \hbar \Delta_0 \hat{\sigma}^{\dag}\hat{\sigma} + \hbar\Delta_{\rm v}\hat{b}^{\dag}\hat{b} + \frac{\hbar g_{\rm S}}{2}\left( \hat{\sigma}\hat{b}^{\dag} + {\rm h.c.} \right).
\end{split}
\end{equation}
Here the detunings are $\Delta_0 = \omega_0 - \tilde{\omega}_{0}$, $\Delta_{\rm v} = \omega_{\rm v} - \tilde{\omega}_{\rm v}$, where the two arbitrary rotating frame's frequencies $\tilde{\omega}_{0}, \tilde{\omega}_{\rm v}$ are chosen to fulfill $\tilde{\omega}_{0}-\tilde{\omega}_{\rm v} = \omega_{\rm S}$.

To observe any coherent interaction with the vibrational states, we need to ensure that the energy scale of the conservative dynamics in Eq. \eqref{eq:ham_JC} overcomes the dissipations described in Sec. \ref{sec:master_eq} by the term $\gamma_{\rm v}$ hence yielding the following condition:
\begin{equation}\label{eq:stokes_strong_condition}
    g_{\rm S} \gtrsim \gamma_{\rm v}.
\end{equation}

In what follows we will discuss a practical example. 
For a DBT molecule the brightest vibrational modes have $\gamma_{\rm v}/(2\pi) \approx 10\,$GHz and a Franck-Condon factor $\eta^2 \approx 0.1$ \cite{clear_phonon-induced_2020, zirkelbach_high-resolution_2022}. Combining Eq. \eqref{eq:g_S_definition} with Eq. \eqref{eq:stokes_strong_condition}, we obtain a condition on the Rabi frequency for the Stokes laser of $\Omega_{\rm S}/(2\pi) \gtrsim 33\,$GHz. 
Using the relation between the Rabi frequency and the electric field intensity in Eq. \eqref{eq:Rabi_power_realtion_general} from App. \ref{app:dimensional_param}, combined with the typical dipole strength of the DBT optical transition $\xi_{0}\approx 0.23$nm \cite{toninelli_single_2021}, we obtain the condition on the required laser intensity $I_{\rm S} \gtrsim 1\,$MW/cm$^2$. 
In a standard experiment \cite{toninelli_single_2021}, this corresponds to the high saturation regime of the off-resonant excitation (also known as incoherent pump, or anti-Stokes electronic transition as we will see in Sec. \ref{sec:anti_stokes}), as highlighted in green in Fig. \ref{fig:2}(b).

\subsection{Jaynes-Cummings vacuum Rabi splitting}

\begin{figure}
    \centering
    \includegraphics[width=\columnwidth]{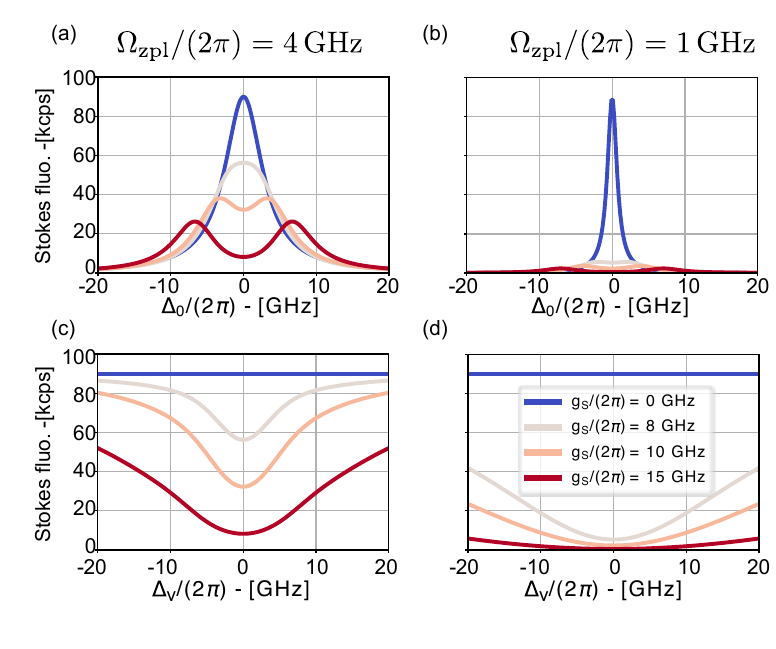}
    \caption{Stokes fluorescence $\Gamma_{\rm S}/(2\pi) = p_{\rm click}\gamma_{0}/(2\pi)\langle{N_{\rm S}\rangle}$ (in kilo-counts per second [kcps]) for the combined Stokes-ZPL drivings, (a) under strong ZPL driving, (b) under weak ZPL driving, (a-b) as a function of ZPL detuning $\Delta_0$, with $\Delta_{\rm v} = \Delta_0$ (sweep on $\omega_{\rm zpl}$). (c) Under strong ZPL driving, (d) under weak ZPL driving, (c-d) as a function of the vibrational detuning $\Delta_{\rm v}$, fixing $\Delta_0=0$ (sweep on $\omega_{\rm S}$). Parameters (all): $\gamma_{0}/(2\pi)=40\,$MHz, $\gamma_{\rm v}/(2\pi)=10\,$GHz, $\eta=0.3$, $p_{\rm click}=0.05$.}
    \label{fig:3}
\end{figure}

As a first case study, we probe the Stokes-JC Hamiltonian in Eq. \eqref{eq:ham_JC} with another laser with frequency $\omega_{\rm zpl}$, resonant with the HOMO-LUMO electronic transition $\omega_0$.
The level scheme is represented in Fig. \ref{fig:2}(c), and the total Hamiltonian reads
\begin{equation}
\begin{split}
    H &= \hbar \Delta_0 \hat{\sigma}^{\dag}\hat{\sigma} + \hbar\Delta_{\rm v}\hat{b}^{\dag}\hat{b} 
    \\
    &+ \frac{\hbar g_{\rm S}}{2}\left( \hat{\sigma}\hat{b}^{\dag} + {\rm h.c.} \right) + \frac{\hbar \Omega_{\rm zpl}}{2}\left( \hat{\sigma}  + {\rm h.c.} \right).
\end{split}
\end{equation}
The detunings are given by $\Delta_0 = \omega_0 - \omega_{\rm zpl}$ and $\Delta_{\rm v}= \omega_{\rm v} - (\omega_{\rm zpl}-\omega_{\rm S})$.

In Fig. \ref{fig:3} the Stokes shifted fluorescence is plotted, $\Gamma_{\rm S}=p_{\rm click}\gamma_{0} \langle{ N_{\rm S} \rangle}$, for the combined ZPL-Stokes drivings as a function of electronic and vibrational detunings, $\Delta_{0}, \Delta_{\rm v}$, and for different Stokes coupling frequencies, $g_{\rm S}$. 
Specifically, in Fig. \ref{fig:3}(a), we show a sweep in the ZPL frequency $\omega_{\rm zpl}$, resulting in a simultaneous sweep of both electronic and vibrational detunings $\Delta_0=\Delta_{\rm v}$. For moderately strong ZPL driving frequency $\Omega_{\rm zpl}/(2\pi)\approx 4\,$GHz, the fluorescence exhibits the characteristic Rabi splitting due to the vibrational dressing of the electronic excitation.
The new two peaks are situated at $\Delta_0 \approx \pm g_{\rm S}$, as recently observed in Ref. \cite{zirkelbach_spectral_2023}, being present also at smaller ZPL power with a sensibly suppressed fluorescence amplitude, as visible in Fig. \ref{fig:3}(b).

On the contrary, keeping $\Delta_0 = 0$ fixed and scanning the Stokes driving frequency $\omega_{\rm S}$ (i.e. scanning the vibrational detuning only $\Delta_{\rm v}$), one observes a large depletion of the fluorescence around $\Delta_{\rm v} = \Delta_0 = 0$.
This behavior is shown in Fig. \ref{fig:3}(c-d).

Although this configuration provides evidence for the electronic-vibrational coherent hybridization through the Rabi splitting, the vibrational state of the molecule remains always very close to its vacuum state, $\langle{ \hat{b}^{\dag}\hat{b} \rangle}\approx 0$ (see App. \ref{app:mean_field_OBE} for a simple semi-classical estimation). As a consequence, the anti-Stokes fluorescence is expected to be almost zero $\langle{N_{\rm AS}\rangle}\approx 0$.

\subsection{Generalized Rabi model and vibrational bi-modality}
\label{sec:anti_stokes}
The second most relevant example is represented in Fig. \ref{fig:2}(d) and is given by combining the Stokes and the anti-Stokes driving terms in Eqs. \eqref{eq:ham_drive_stokes}-\eqref{eq:ham_drive_antistokes}, fixing to zero the ZPL-drive $\Omega_{\rm zpl}=0$.
Defining the detunings $\Delta_0 = \omega_0 - (\omega_{\rm AS}+\omega_{\rm S})/2$ and $\Delta_{\rm v}= \omega_{\rm v} - (\omega_{\rm AS}-\omega_{\rm S})/2$, the system is expressed in the form of a generalized Rabi Hamiltonian
\begin{equation}\label{eq:ham_generalised_Rabi}
\begin{split}
    H_{\rm gRabi} & = \hbar \Delta_0 \hat{\sigma}^{\dag}\hat{\sigma} + \hbar\Delta_{\rm v}\hat{b}^{\dag}\hat{b} 
    \\
    &+ \frac{\hbar g_{\rm S}}{2}\left( \hat{\sigma}\hat{b}^{\dag} + {\rm h.c.} \right) + \frac{\hbar g_{\rm AS}}{2}\left( \hat{\sigma}^{\dag} \hat{b}^{\dag} + {\rm h.c.} \right).
\end{split}
\end{equation}

The anti-Stokes drive introduces an anti-Jaynes-Cummings interaction in Eq. \eqref{eq:ham_generalised_Rabi}, which falls in the category of two-mode squeezing parametric amplifiers \cite{walls_quantum_2008}. 
Considering first $g_{\rm S}=0$, due to the large mismatching between vibrational and optical dissipation $\gamma_{\rm v} \gg \gamma_{0}$ the vibrational mode is almost empty $\langle{\hat{b}^{\dag}\hat{b}\rangle}\approx 0$.
Having a very dissipative mode without any relevant dynamics, we can consider it as a virtual level tracing it away by performing a Born-Markov approximation. As a result, we obtain an effective Lindbladian incoherent pumping term (see App. \ref{app:sat_curve}) that accounts for what is usually termed in experiments as off-resonant pumping \cite{duquennoy_real-time_2022}.

When the Stokes coupling is also non-zero, $g_{\rm S}\neq 0$, and sufficiently strong with respect to the dissipation, $g_{\rm S} \simeq \gamma_{\rm v}$, the vibrational mode is populated by a finite amount of excitations, and the vibrational dynamics cannot be traced away.
The mechanism works as follows: the anti-Stokes laser pumps up the electronic state together with a quantum of vibration; the strong Jaynes-Cummings interaction converts the electronic excitation into a new vibrational quantum. The process is then repeated, starting from a vibrational state that already contains two quanta of excitations and thus leading to a net increase of the vibrational population.

\begin{figure}
    \centering
    \includegraphics[width=\columnwidth]{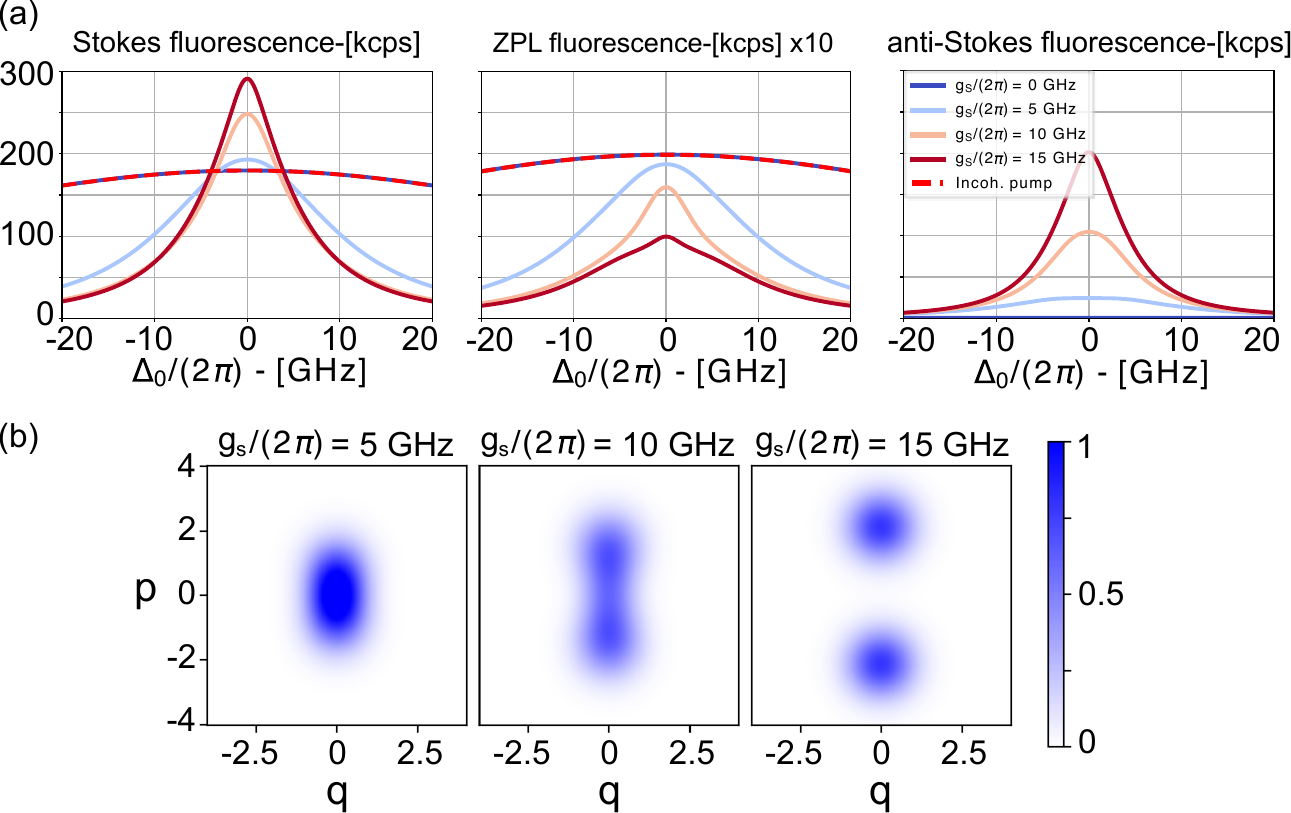}
    \caption{(a) Stokes, ZPL and anti-Stokes fluorescence (in kcounts per second [kcps]) as a function of $\Delta_0$, with $\Delta_{\rm v}=\Delta_0$ (sweep on $\omega_{\rm AS}$). 
    (b) Normalized Wigner function $W_{\rm v}(q,p)$ of the vibrational steady-state at $\Delta_{\rm v}=\Delta_0=0$ (other parameters are the same as in (a)).
    Parameters: $\gamma_{0}/(2\pi)=40\,$MHz, $\gamma_{\rm v}/(2\pi)=10\,$GHz, $g_{\rm AS}/(2\pi)=15\,$GHz, $\eta=0.3$, $p_{\rm click}=0.05$. }
    \label{fig:4}
\end{figure}

In Fig. \ref{fig:4}(a), we look at the emitted Stokes, ZPL, and anti-Stokes fluorescence as a function of the anti-Stokes resonance frequency $\omega_{\rm AS}$ and for various values of the Stokes coupling $g_{\rm S}$. The anti-Stokes coupling is fixed to $g_{\rm AS}/(2\pi)=15\,$GHz such that the system is deep into the saturation regime.

At $g_{\rm S} = 0$ the system is well described by the already mentioned incoherent pumping term in Eq. \eqref{eq:Gamma_plus_up_rate}-\eqref{eq:rho_ee_incoherent_drive_anal}, and the fluorescence is strong both in the ZPL and the Stokes channels (Fig. \ref{fig:4}(a) left and central panel), whereas the anti-Stokes emission is almost zero (Fig. \ref{fig:4}(a) right panel).
When the Stokes drive is switched on, $g_{\rm S} \neq 0$, the Stokes emission increases, sensibly exceeding the saturation value described by the incoherent pump, while the ZPL fluorescence is reduced. 
Differently from the pure JC model discussed in the previous subsection, here we observe sizable emission also at the anti-Stokes frequency, Fig. \ref{fig:4}(a-right panel), indicating the accumulation of a finite population in the vibrational mode, $\langle{\hat{b}^{\dag} \hat{b} \rangle} \neq 0$.
To better grasp the specific quantum state which is populated, in Fig. \ref{fig:4}(b) we plot the Wigner function of the vibrational steady-state $W_{\rm v}(q,p)$ as a function of the quadrature variables (see Ref. \cite{johansson_qutip_2013}, or the online QuTip docs for the detailed definition).
When both Stokes and anti-Stokes couplings exceed the vibrational dissipation, $g_{\rm S}, g_{\rm AS} \geq \gamma_{\rm v}$, the steady state is characterized by a bimodal vibrational Wigner function.
Similarly to photonic non-linear Kerr oscillators under two-photon drive \cite{minganti_exact_2016}, the $\mathbb{Z}_2$ symmetry, $\hat{\sigma}\mapsto -\hat{\sigma}$ and $\hat{b}\mapsto-\hat{b}$, of the dissipative Rabi model \cite{hwang_dissipative_2018} ensure a symmetric steady state which, as we will see in the following, is a statistical mixture of vibrational cat states \cite{bartolo_homodyne_2017}.

It is worth pointing out that, under the condition of equal Stokes and anti-Stokes drivings amplitudes, $g_{\rm S} = g_{\rm AS} = g$, the Hamiltonian in Eq. \eqref{eq:ham_generalised_Rabi} becomes the symmetric quantum Rabi model. 
This is one of the most paradigmatic models to study light-matter interactions at the quantum level \cite{braak_semi-classical_2016,de_bernardis_tutorial_2024, pedernales_quantum_2015, lv_quantum_2018, gerritsma_quantum_2010, hwang_quantum_2015,hwang_dissipative_2018, cai_observation_2021}, making molecules a valuable quantum simulation platform.

\section{Vibrational cat states and quantum jumps}
\label{sec:blinking}
While the anti-Stokes emission is a clear signature of a finite vibrational population, observing the bimodality of the Wigner function in Fig. \ref{fig:4}(b) is a subtle issue.
As remarked in Ref. \cite{bartolo_homodyne_2017}, the $\mathbb{Z}_2$ symmetry of the generalized Rabi model in Eq. \eqref{eq:ham_generalised_Rabi}, preserved by the molecule's main dissipation channels described in Sec. \ref{sec:master_eq}, prevents to observe the bimodality in any quantity averaged over the steady-state density matrix.
To circumvent this limitation, one can use the information acquired by continuously monitoring the system, unraveling the time evolution of the density matrix, and analyzing the behavior of each single \emph{quantum trajectory} \cite{molmer_monte_1993,gardiner_physics_2015, minganti_open_2024}.

To this end, we make use of the Monte-Carlo algorithm implemented in QuTip \cite{johansson_qutip_2013} to solve the Stochastic Schr\"odinger equation related to the dissipative dynamics described in Sec. \ref{sec:master_eq}, where the Hamiltonian is given by the symmetric Rabi model in Eq. \eqref{eq:ham_generalised_Rabi} with $g_{\rm S} = g_{\rm AS} = g$. 
We use $|\Psi(t)\rangle$ to denote a single unraveled trajectory, for which the steady-state density matrix is given by averaging over its multiple realizations $\hat{\rho}_{ss} \sim \sum_{\Psi} |\Psi(t)\rangle \langle \Psi(t) |$.

\begin{figure}
    \centering
    \includegraphics[width=\columnwidth]{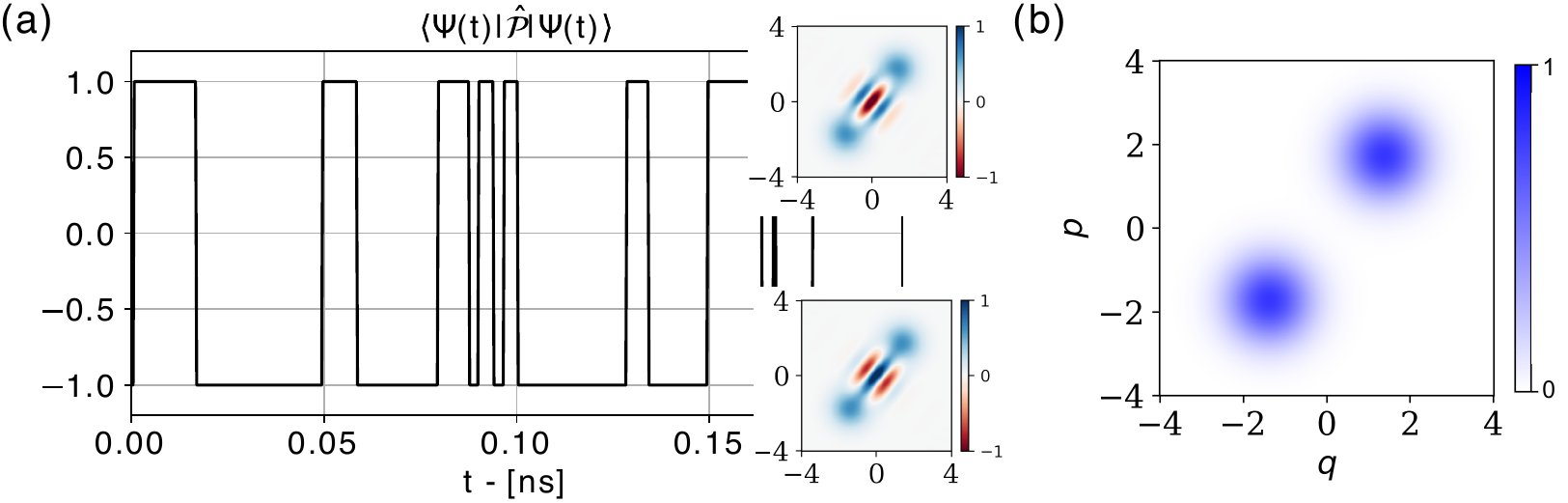}
    \caption{(a) Time evolution of the expectation value of the parity operator $\langle{\Psi(t)| \hat{\mathcal{P}} | \Psi(t) \rangle}$ over a single trajectory $| \Psi(t) \rangle$. Insets: excited-state projected Wigner function (normalized) of the instantaneous state from a single trajectory, as defined in the main text. The upper one is for a state with parity $\braket{\hat{\mathcal{P}}} = 1$, while the lower one is for a state with parity $\braket{\hat{\mathcal{P}}} = -1$. (b) Wigner function associated with the trajectory-averaged steady-state density matrix (normalized). Notice that the presence of finite $\Delta_{\rm v}\neq 0$ rotates the bi-modal distribution in the phase space.  Parameters: $\Delta_{0}/(2\pi)=0$, $\Delta_{\rm v}/(2\pi)=4\,$GHz, $g/(2\pi)=20\,$GHz, $\eta = 0.3$, $\gamma_{\rm v}/(2\pi)=10\,$GHz, $\gamma_{0}/(2\pi)=0.04\,$GHz, vibrational numerical cutoff $N_{\rm v-cutoff}=50$.}
    \label{fig:5}
\end{figure}

It turns out that the bimodality is visible as a jumping signal (telegraph signal \cite{blatt_quantum_1988}) in the time evolution of the $\mathbb{Z}_2$ parity operator $\hat{\mathcal{P}}=\exp\left[ i\pi(\hat{\sigma}^{\dag}\hat{\sigma} + \hat{b}^{\dag}\hat{b}) \right]$, as for the two-photon Kerr resonator \cite{minganti_exact_2016,bartolo_homodyne_2017}. 
As shown in Fig. \ref{fig:5}(a), starting from an arbitrary initial state, after sufficient long time the expectation value of $\langle \Psi(t)|\hat{\mathcal{P}}|\Psi(t)\rangle$ enters in a steady random jump dynamics between $\pm 1$ values.

An intuitive understanding is provided by considering the two opposite parity states
\begin{equation}
     |\Psi_{\pm}\rangle =  \frac{1}{\sqrt{2}}\left[ |{\rm cat}_{\pm}\rangle |g\rangle  + |{\rm cat}_{\mp}\rangle  |e\rangle \right].
\end{equation}
Here $|{\rm cat}_{\pm}\rangle = (|\beta\rangle \pm |-\beta\rangle)/\sqrt{\mathcal{N}_{\pm}}$ is a cat state made of vibrational coherent states with amplitude $\beta$, and $\mathcal{N}_{\pm} = 2\pm 2\exp [-2|\beta|^2]$, where the coherent state amplitude scales with the ratio between the Rabi coupling and the vibrational decay $\beta\sim g/\gamma_{\rm v}$ \cite{hwang_dissipative_2018}. 

$|\Psi_{\pm}\rangle$ are a good approximation of the two lowest eigenstates of the quantum Rabi model in the ultra-strong coupling limit $g\gg \Delta_0, \Delta_{\rm v}$ \cite{de_bernardis_tutorial_2024} and they almost do not evolve under the action of the Hamiltonian.
On the other side, they transform into each other under the effect of dissipation: suppose, for instance, that the system is in the state $|\Psi_+\rangle$. If a vibrational quanta is absorbed by the environment, the state jumps to the new state $\hat{b}|\Psi_+\rangle \sim |\Psi_-\rangle$.
We then expect that the steady-state density matrix is, to good approximation, a statistical mixture of these states.
As a consequence, the quantum jumps caused by the spontaneous emission of a vibrational quantum make the system jump between states of opposite parity, creating the telegraph signal in the parity expectation in Fig. \ref{fig:5}(a).
Notice that the states $|\Psi_{\pm}\rangle$ have the same vibrational and electronic population, and as a consequence, no blinking is visible in the fluorescence.

A simple verification of our reasoning is given by considering the density matrix of the instantaneous single trajectory state projected on the excited electronic state $\hat{\rho}_{\rm proj}(t) = \hat{\sigma}^{\dag}\hat{\sigma}| \Psi(t) \rangle \langle\Psi(t)| \hat{\sigma}^{\dag}\hat{\sigma}$.
Looking at its Wigner function, we recover the shape of opposite parity cat states \cite{deleglise_reconstruction_2008,minganti_exact_2016}, as shown in the insets of Fig. \ref{fig:5}(a).
Instead, when we average the state over many trajectories, the steady-state projected Wigner function remains bi-modal but loses all the quantum negativity as shown in Fig. \ref{fig:5}(b).

While the telegraph signal in the parity is a smoking gun evidence for the vibrational bi-modality, this operator is hardly measurable in a realistic setting since it requires measuring the electronic and vibrational population and all their combined correlations. 
Moreover, as shown in Fig. \ref{fig:5}(a), the jumps occur on the fast timescales of the vibrational dissipation $t_{\rm jump}\sim 1/\gamma_{\rm v} < 0.1$ ns, much faster than the typical detector resolution.

A different strategy is instead to explicitly break the symmetry with a bias laser and then observe the blinking in the photon counting signal. 
This occurs on slower time scales, set by the interplay between the asymmetric bias and the symmetric dissipative dynamics.
Clearly, the system is then driven to a different steady state, and the jumps happen between slightly different states. However, if the bias is not too strong, some features of the symmetric state will remain, thus providing a method to, at least, prove its existence.

To this aim, we consider the second order in the polaron expansion due to Eq. \eqref{eq:displacement_normal_order_expansion} by adding a new laser with frequency $\omega_{\rm AS^{\mathbi{2}}} = (3\omega_{\rm AS} - \omega_{\rm S})/2$, which is resonant with the two-vibrations transition, as illustrated in Fig. \ref{fig:2}(e).
In the same way as explained for the ZPL, Stokes, and anti-Stokes Hamiltonians in Sec. \ref{sec:polaron_dressing}, we add the new Hamiltonian term to our toolbox
\begin{equation}
    H_{\rm AS^{\mathbi{2}}} = \frac{\hbar g_{\rm AS^{\mathbi{2}}}}{2}\left( \hat{\sigma}\hat{b}^2 + {\rm h.c.} \right),
\end{equation}
where $g_{\rm AS^{\mathbi{2}}} = \eta^2\Omega_{\rm AS^{\mathbi{2}}}/2$ is the \emph{anti-Stokes}$^{\mathbi{2}}$ two-vibrations coupling constant, proportional to the respective laser Rabi frequency $\Omega_{\rm AS^{\mathbi{2}}}$. Notice that the $\sim \eta^2$ suppression makes this term quite small but still relevant for our purposes.
Indeed, in complete analogy with the two-photon Rabi model \cite{felicetti_two-photon_2018}, this term is invariant under the two transformations $\hat{b}\mapsto -\hat{b}$, $\hat{\sigma}\mapsto\hat{\sigma}$ and $\hat{b}\mapsto i\hat{b}$, $\hat{\sigma}\mapsto -\hat{\sigma}$, but does not respect the original $\mathbb{Z}_2$ symmetry of the Rabi model.

\begin{figure}
    \centering
    \includegraphics[width=\columnwidth]{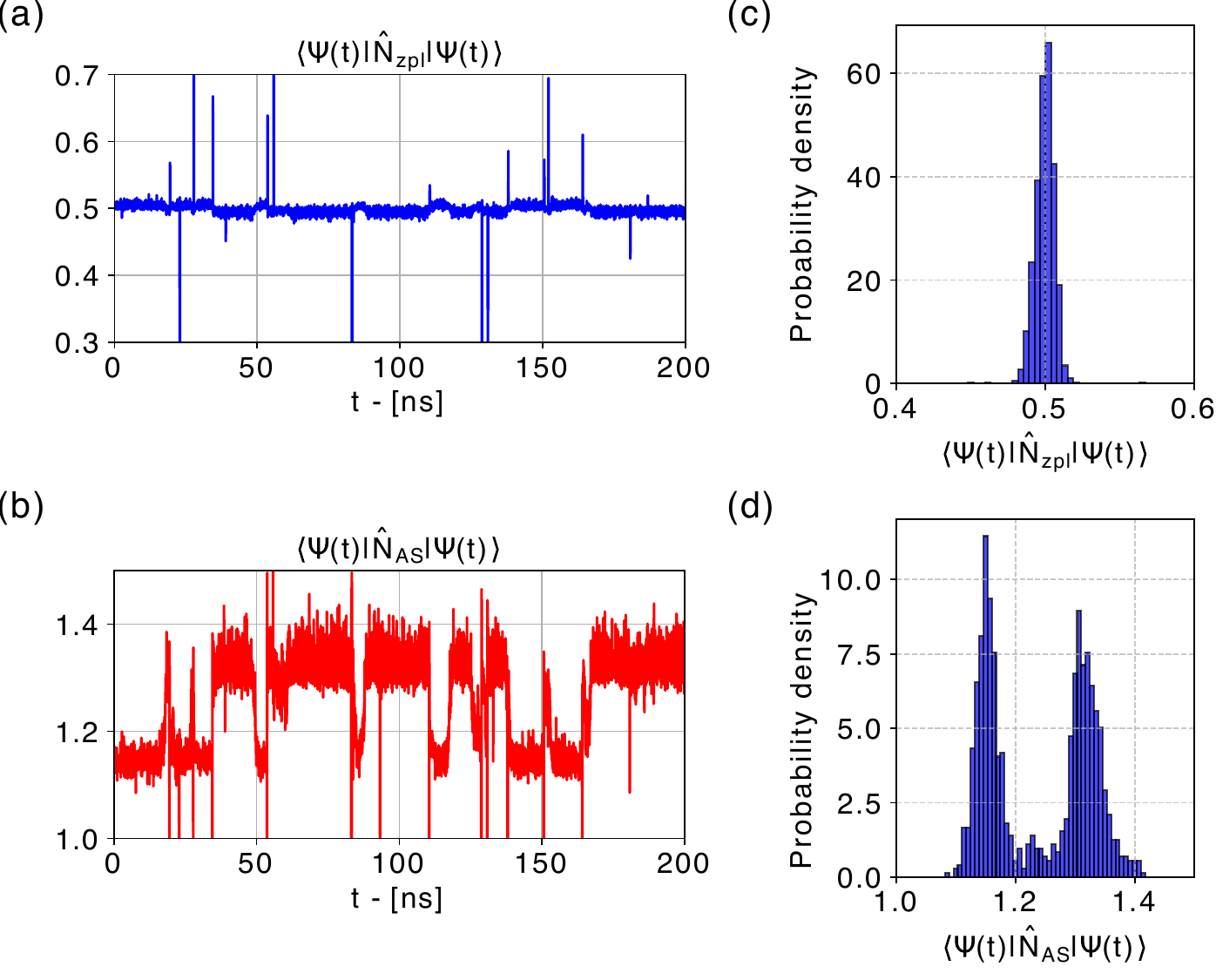}
    \caption{(a) ZPL single trajectory photon number $\langle{\Psi(t)| \hat{N}_{\rm zpl} | \Psi(t) \rangle}$  with the bias anti-Stokes$^2$ driving term. (b) Anti-Stokes single trajectory photon number $\langle{\Psi(t)| \hat{N}_{\rm AS} | \Psi(t) \rangle}$  with the bias anti-Stokes$^2$ driving term.
    (c) Histogram of $\langle{\Psi(t)| \hat{N}_{\rm zpl} | \Psi(t) \rangle}$ (d) Histogram of $\langle{\Psi(t)| \hat{N}_{\rm AS} | \Psi(t) \rangle}$. Both (a-b) are made taking $N_{\rm traj}= 1000$ trajectories, with the same fixed time $t$, taken after reaching the steady-state. 
    Parameters: $g_{\rm AS^{\mathbi{2}}}/(2\pi)= 1.5\,$GHz; the system is initialized a low-energy random state and evolved until reaching the steady state.
    Other parameters: same as Fig. \ref{fig:5}(c-d) }
    \label{fig:5}
\end{figure}

The total Hamiltonian of the system is now given by $H_{\rm gRabi}+H_{\rm AS^{\mathbi{2}}}$ under the assumption that $g_{\rm AS^{\mathbi{2}}} < g_{\rm S}=g_{\rm AS} = g$, where the equality between the Stokes and anti-Stokes drivings is not necessary, but simplifies the discussion.
In Fig. \ref{fig:5}(a-b) we show the instantaneous photon number $\langle{\Psi(t)|\hat{N}_{\rm zpl}|\Psi(t)\rangle}$, $\langle{\Psi(t)|\hat{N}_{\rm AS}|\Psi(t)\rangle}$ for a single unraveled trajectory $|\Psi(t)\rangle$. 
The presence of the bias anti-Stokes$^{\mathbi{2}}$ driving term induces a blinking in the photon counting by breaking the symmetry in the bi-modal vibrational steady-state. The blinking occurs now on a much longer timescale, that can be observed with the current technology.
This phenomenon is particularly visible in the anti-Stokes photon number, where the jumps have a size comparable with the average signal.

Repeating the algorithm over many trajectories, we can collect the blinking values in a histogram to visualize the probability density of having a certain photon number.
In Fig. \ref{fig:5}(c) we show what is obtained for the ZPL photon number $\langle{\Psi(t)|\hat{N}_{\rm zpl}|\Psi(t)\rangle}$. The very small jumps partially visible in Fig. \ref{fig:5}(a) give rise to only a small broadening in the histogram probability density. However, the ZPL photon number does not exhibit any bimodality.
On contrary, the histogram of the anti-Stokes photon number $\langle{\Psi(t)|\hat{N}_{\rm AS}|\Psi(t)\rangle}$, shown in Fig. \ref{fig:5}(d), is clearly bi-modal. As a consequence, the anti-Stokes fluorescence appears to be a good witness of the bi-modality.

\begin{figure}
    \centering
    \includegraphics[width=\columnwidth]{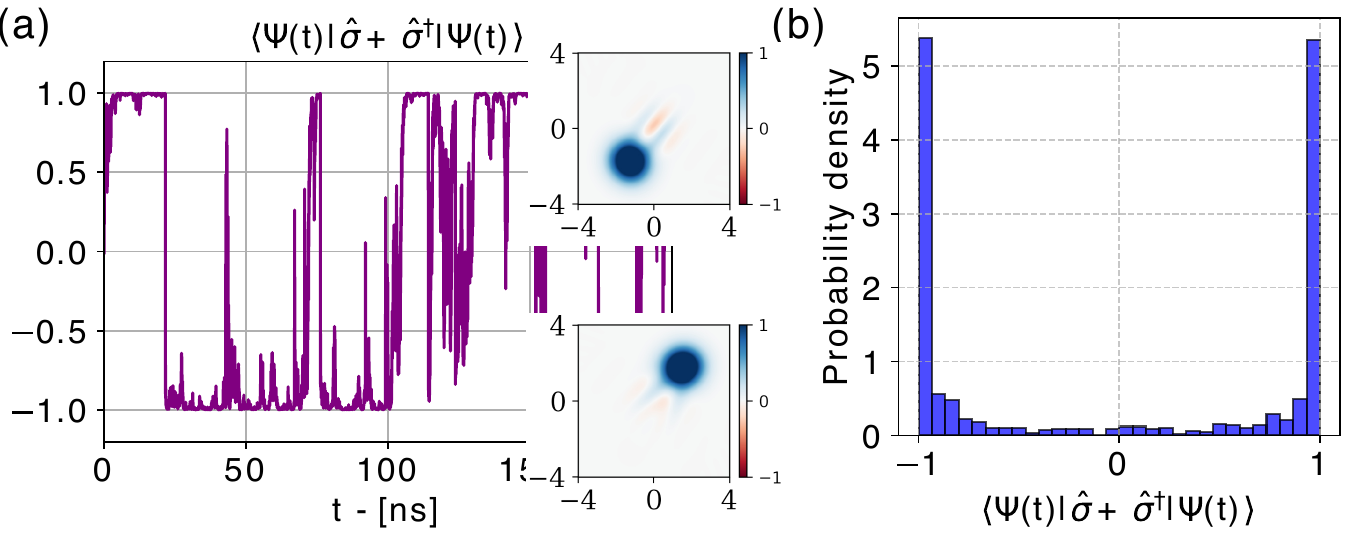}
    \caption{(a) Time evolution of the expectation value of $\langle{\Psi(t)| \hat{N}_{\rm zpl} | \Psi(t) \rangle}$ over a single trajectory $| \Psi(t) \rangle$. Insets: excited-state projected Wigner function (normalized) of the instantaneous state from a single trajectory, as defined in the main text. The upper one is for a state with $\braket{\hat{\sigma} + \hat{\sigma}^{\dag}} = 1$, while the lower one is for a state with parity $\braket{\hat{\sigma} + \hat{\sigma}^{\dag}} = -1$.  (b) Histogram of $\langle{\Psi(t)| \hat{\sigma} + \hat{\sigma}^{\dag} | \Psi(t) \rangle}$ taken over $N_{\rm traj}= 1000$ trajectories at the same fixed time $t$ after reaching the steady-state. Parameters: same as Fig. \ref{fig:5}(c-d) }
    \label{fig:7}
\end{figure}

The fact that the vibrational bi-modality does not show up in all the observables reminds Ref. \cite{bartolo_homodyne_2017} where switching from photon counting to homodyne detection is crucial to observe the jumps.
Following this reasoning, we provide another striking example by looking at the expectation value of the ZPL electric field operator $\hat{E}_{\rm zpl}\sim \hat{\sigma} + \hat{\sigma}^{\dag}$, which is the main observable in the homodyne detection.
Constrained by the symmetry of the Rabi model, at zero anti-Stokes$^{\mathbi{2}}$ bias, $g_{\rm AS^{\mathbi{2}}}=0$, this quantity is always exactly zero, $\langle{\Psi(t)| \hat{\sigma} + \hat{\sigma}^{\dag} | \Psi(t) \rangle}|_{g_{\rm AS^{\mathbi{2}}}=0} = 0$.
In Fig. \ref{fig:7}(a) we show the time evolution of $\langle{\Psi(t)| \hat{\sigma} + \hat{\sigma}^{\dag} | \Psi(t) \rangle}$ over a single trajectory, for $g_{\rm AS^{\mathbi{2}}}/(2\pi)=1.5\,$GHz, which exhibits a telegraph signal jumping between $\pm 1$ values.
As in the previous cases, in Fig. \ref{fig:7}(b), we plot the probability density histogram by collecting multiple trajectories, which shows sharp bi-modality over the extremal values.
In the insets of Fig. \ref{fig:7}(a) we show the excited electronic state projected Wigner function, associated to the istantaneous single trajectory state $\hat{\rho}_{\rm proj}(t)$ for the extremal values $\langle{\Psi(t)| \hat{\sigma} + \hat{\sigma}^{\dag} | \Psi(t) \rangle}=\pm 1$. While the bias still preserves a small amount of negativity, the cat-state structure is mostly gone due to the breaking of the symmetry.
On the contrary, when averaged over many trajectories, one recovers a Wigner function almost identical to Fig. \ref{fig:5}(b).
As a disclaimer, we must say that a full discussion and characterization of homodyne detection for these cryogenic molecular setups would be too involved and far beyond the scope of the current paper. 
Here we limit ourselves to acknowledging the possibility of observing the vibrational bi-modality through various quantities by explicitly breaking the symmetry of the system.

\section{THz-to-optical transducer}
\label{sec:trans}
We finally focus on the consequences of having a finite THz driving $\Omega_{\rm THz} \neq 0$ directly applied to the vibrational dipole transition.

An interesting perspective discussed in Refs. \cite{roelli_molecular_2016,roelli_molecular_2020} and realized experimentally in Refs. \cite{chen_continuous-wave_2021, whaley-mayda_fluorescence-encoded_2021, chikkaraddy_single-molecule_2023, xomalis_detecting_2021} is to exploit the interaction between vibrational and electronic degrees of freedom in molecules to implement an IR-to-optical transducer.
This idea is particularly promising when generalized to large fluorescent organic molecules with similar features as the PAHs.
Having a large quantum yield and low levels of decoherence (both in the ZPL and the vibrational side bands), they operate fully in the quantum regime.
Their low-frequency vibrational modes are typically in the few-THz range, coupled with sizable Franck-Condon factors, $\eta^2\sim 0.01-0.1$, to the electronic transition at $350-750\,$THz \cite{gurlek_small_2024}.
The importance of such a transduction device is immediately understood by considering it as the core of a single THz-photon detector \cite{todorov_thz_2024, ukirade_review_2025}.

The framework of hybrid quantum states engineering, discussed in the previous sections, is also very helpful to develop the basics of the general quantum theory of such a device. 
All the relevant frequency scales emerge clearly, allowing the establishment of the general conditions and figure of merits to achieve the optimal operational regime \cite{lauk_perspectives_2020}.

We consider the Stokes JC Hamiltonian discussed in Eq. \eqref{eq:ham_JC} provided with a THz input, directly driving the vibrational mode
\begin{equation}
\begin{split}
    H_{\rm trans} = & \hbar \Delta_0 \hat{\sigma}^{\dag}\hat{\sigma} + \hbar \Delta_{\rm v} \hat{b}^{\dag}\hat{b} + \frac{\hbar g_{\rm S}}{2}\left(\hat{\sigma} \hat{b}^{\dag} + {\rm h.c.} \right)
    \\
    & + \frac{\hbar \Omega_{\rm THz}}{2}\left( \hat{b} + {\rm h.c.} \right).
\end{split}
\end{equation}
where $\Delta_0= \omega_0 - (\omega_{\rm S} - \omega_{\rm THz})$, $\Delta_{\rm v} = \omega_{\rm v} - \omega_{\rm THz}$.

Under the Stokes driving only, the fluorescence is completely due to the vibrational population, implying that the rate of photon transduction from THz to optical frequency is here given by the total emission rate
\begin{equation}
    \Gamma_{\rm trans} = \gamma_{0} \langle{ \hat{\sigma}^{\dag}\hat{\sigma} \rangle}.
\end{equation}
In this work, we mainly focus on the linear regime, where
\begin{enumerate}
    \item the transducer is linear in the input THz intensity, $I_{\rm THz}$, so its transduction rate scales as $\Gamma_{\rm trans} \sim \Omega_{\rm THz}^2$;
    \item the optical response to a THz photon is maximized.
\end{enumerate}

Following the semi-classical steady-state equations in App. \ref{app:mean_field_OBE}, we can derive an expression that follows the requirement of linearity, reading
\begin{equation}
    \Gamma_{\rm trans} \approx  \chi_{\rm trans} \frac{\Omega_{\rm THz}^2}{\gamma_{0}},
\end{equation}
where the linearized resonant transducer susceptibility is
\begin{equation}\label{eq:trans_susceptibility}
    \chi_{\rm trans} = 2n_{\rm S}\frac{\mathcal{C}_{\rm S}^2}{(1+\mathcal{C}_{\rm S})^2}.
\end{equation}
Here, the THz driving is assumed to be resonant with the vibrational mode $\Delta_{\rm v} = 0$.
Moreover, in complete analogy with the field of cavity QED \cite{kimble_strong_1998}, we have introduced the Stokes cooperativity and saturation number
\begin{align}\label{eq:coop_satnum_stokes}
    \mathcal{C}_{\rm S} = \frac{g_{\rm S}^2}{\gamma_{\rm v} \gamma_{0}}, && n_{{\rm S}}=  \frac{4\Delta_0^2 + \gamma_{0}^2}{2g_{\rm S}^2}.
\end{align}
These quantities are the relevant combinations of parameters characterizing the interplay between electronic and vibrational transitions.

At first sight, it is clear from Eq. \eqref{eq:trans_susceptibility} that, for the transducer to function efficiently, the Stokes cooperativity should be maximized, just like in the strong coupling regime of cavity QED \cite{kimble_strong_1998}. 
Instead, opposite to the strong coupling regime, the saturation number should also be maximized. 
This leads to a competition since they both depend on the Stokes coupling $g_{\rm S}$ with inverse proportionality between each other.
We then deduce that the working regime for a good transducer is different from the strong coupling regime of cavity QED, and it can be understood as follows:
The molecule has to convert efficiently the vibrational excitation to an electronic one, quicker than the vibrational dissipation and thus requiring strong Stokes coupling. 
However, once the conversion is done, it needs to quickly flush away the excitation as an optical photon before it can be reconverted into a vibration, requiring a Stokes coupling weaker than the radiative (electronic) decay rate.

\begin{figure}
    \centering
    \includegraphics[width=\columnwidth]{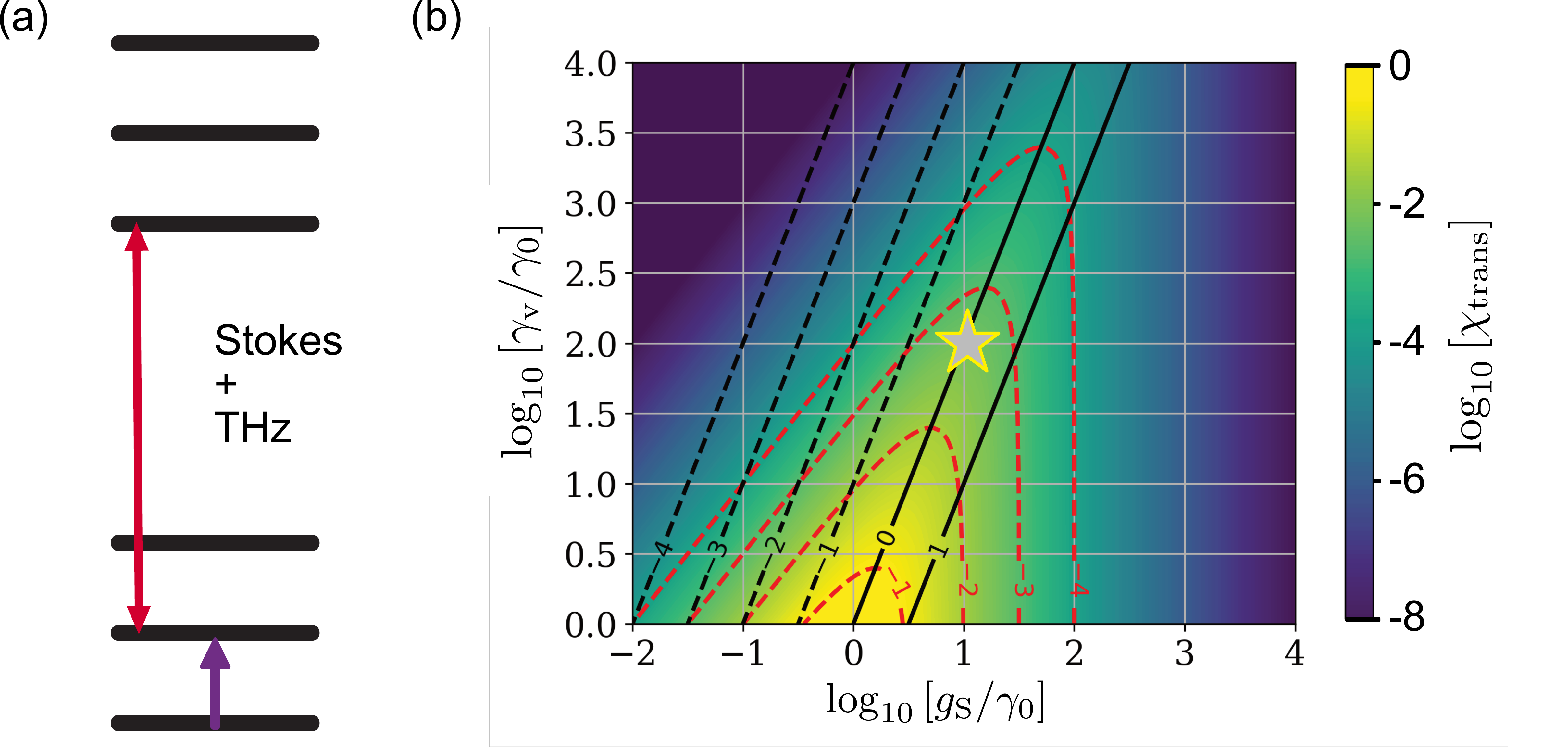}
    \caption{(a) Level scheme representation of the combined Stokes and THz drive. The small purple arrow represents the THz driving. (b)The transducer susceptibility as a function of the normalized Stokes coupling $g_{\rm S}/\gamma_{0}$ and the normalized vibrational dissipation rate $\gamma_{\rm v}/\gamma_{0}$. The black lines are isolines at fixed cooperativity value $\log_{10}[\mathcal{C}_{\rm S}] ={\rm const}$. The red lines are the isolines of $\chi_{\rm trans}$ in logscale. Solid lines are for positive values and dashed for negative ones. The star marks the typical value for a PAH-like molecule.
    Other parameters: $\Delta_0 = \Delta_{\rm v} = 0$.}
    \label{fig:8}
\end{figure}

These conditions altogether lead to a non-trivial scaling of the transduction susceptibility, which is visualized in Fig. \ref{fig:8}. 
Here we plot Eq. \eqref{eq:trans_susceptibility} as a function of the Stokes coupling $g_{\rm S}$ and the vibrational linewidth $\gamma_{\rm v}$ normalized on the electronic decay rate $\gamma_{0}$ which is kept fixed. The isolines of constant cooperativity are plotted as black lines (in log-scale).
Since the cooperativity and the saturation number are not independent, there exists a best trade-off, which appears at $\mathcal{C}_{\rm S} = 1$.
This condition is thus interpreted as an optimal impedance matching condition in the vibrational-electronic excitation transduction.

The optimal case for a candidate molecule with properties similar to those of DBT is marked in Fig. \ref{fig:8} by a star, in the region where $\chi_{\rm trans} \sim 10^{-3}-10^{-2}$.
Assuming a THz-drive Rabi frequency of $\Omega_{\rm THz}/(2\pi)= 100\,$MHz (this value will be clarified in the following section), $\gamma_{0}/(2\pi) = 40\,$MHz and the pessimistic value of $\chi_{\rm trans}=10^{-3}$ we obtain an emission rate of $\Gamma_{\rm trans}/(2\pi) \approx 250\,$kHz. 
With a detector click probability $p_{\rm click}=0.05$, we would observe a fluorescence of $\sim 12.5\,$kcps (kilo-count per second), very similar to what was recently observed for IR-optical transduction at room temperature in Ref. \cite{chikkaraddy_single-molecule_2023}.
We remark that our analysis is very general and can be thus adapted to all kinds of IR-optical transduction phenomena with molecules.

Overall, this analysis states that a good fluorescent molecular transducer requires a large quantum yield (i.e., large optical dipole transition) and narrow vibrational transitions.
The amplitude of the Stokes driving is instead fixed by maximizing Eq. \eqref{eq:trans_susceptibility} (or looking at Fig. \ref{fig:8}), which typically corresponds to work in a high-saturation regime.

\section{Transducer's input implementations}
\label{sec:implementation_THZ}
While in the previous section, we provided the general description for the transducer output, here we develop a basic design for its input.
This stands to quantify the amount of THz radiation that can reach the molecule and what is the amount that is absorbed.
In this way, we can provide a concrete estimation of $\Omega_{\rm THz}$ that can be obtained in real experiments.

Since the Rabi frequency is just given by the energy of a dipole in a electric field $\hbar \Omega_{\rm THz} = e\xi_{\rm v}E_{\rm THz}$, the problem is split in two: estimate the vibrational transition dipole $\xi_{\rm v}$ and the THz RMS electric field $E_{\rm THz}$ hitting the molecule.
Here $\xi_{\rm v} = \langle{0_{\rm v}| \hat{\xi}_{\rm v}  | 1_{\rm v}\rangle}$ is the matrix element of the dipole operator $\hat{\xi}_{\rm v}$ associated to the vibrational mode, $|0_{\rm v}\rangle$ is the vibrational vacuum and $ | 1_{\rm v}\rangle = \hat{b}^{\dag} | 0_{\rm v}\rangle$ is the first vibrational excitation.

\subsection{Vibrational transition dipole and selection rule}
Evaluating $\xi_{\rm v}$ for a specific vibrational mode from first principles is not an easy task, relying on complex DFT simulations \cite{kostjukova_vibronic_2021}, and is far beyond the aim of our current work.
However, we can still have an order of magnitude estimation based on the available measurements. 
Following Ref. \cite{tretyakov_maximizing_2014,jeannin_absorption_2020}, we consider the fundamental relation between resonant absorption cross section and radiative decay rate as
\begin{equation}
    \frac{P_{\rm abs}}{I}=\sigma_{\rm abs}^{\rm v} = \frac{3\lambda_{\rm v}^2}{8\pi} \frac{\gamma_{\rm v} \gamma_{\rm rad}^{\rm v}}{(\gamma_{\rm v} + \gamma_{\rm rad}^{\rm v})^2},
\end{equation}
where $P_{\rm abs}$ is the absorbed power, $I$ is the incident radiation intensity, $\lambda_{\rm v} = 2\pi c/\omega_{\rm v}$ is the vibrational wave length, and the radiative rate is linked to the transition dipole strength through \cite{jackson_classical_2013, tretyakov_maximizing_2014}
\begin{equation}
    \gamma_{\rm rad}^{\rm v} = \frac{4}{3}\alpha_{\rm fs}   \frac{\omega_{\rm v}^3}{c^2} \xi_{\rm v}^2.
\end{equation}
Here, $c$ is the speed of light, and $\alpha_{\rm fs} = e^2/(4\pi\epsilon_0 \hbar c) \approx 1/137$ is the fine-structure constant.

Having that the vibrational non radiative decay is much bigger than the radiative one $\gamma_{\rm v} \gg \gamma_{\rm rad}^{\rm v}$, we can derive the vibrational transition dipole as a function of the absorption cross section as in Ref. \cite{roelli_molecular_2020}
\begin{equation}\label{eq:dipole_cross_section}
    \xi_{\rm v} \approx \sqrt{\frac{\sigma_{\rm abs}^{\rm v}}{4\pi \alpha_{\rm fs}\mathcal{Q}_{\rm v}}}.
\end{equation}
Here we introduced the vibrational quality factor as $\mathcal{Q}_{\rm v} = \omega_{\rm v}/\gamma_{\rm v}$.
For instance, considering the IR-optical transduction experiment in Ref. \cite{chikkaraddy_single-molecule_2023}, it was estimated $\sigma_{\rm abs}^{\rm v} \sim 10^{-5}\,$nm$^{2}$. Assuming a vibrational quality factor $\mathcal{Q}_{\rm v} \sim 10^2$, we obtain $\xi_{\rm v} \sim 10^{-3}\,$nm.
In other IR-active molecules, like in Ref. \cite{shalabney_coherent_2015}, it can arrive up to $\xi_{\rm v}\sim 10^{-2}\,$nm. 
Based on evidence of similar strong absorbance, we expect the estimations for IR active molecules to hold also for molecules with modes active down to the THz range, like PAH molecules \cite{han_terahertz_2004, cataldo_far_2013}. 
From here on, we take this range as a reference.

While PAH molecules like DBT seem to have optimal parameters to implement the THz-optical transducer, they face a fundamental limitation imposed by their high level of symmetry.
For every centrosymmetric molecule, the "IR-Raman" selection rule imposes that $\eta \neq 0 \implies \xi_{\rm v} = 0$ (and vice versa) \cite{di_bartolo_optical_2010}, making it not possible to have non-zero Stokes coupling together with non-zero THz Rabi frequency.
This bound can be in principle circumvented by engineering specific non-symmetric molecules that preserve the condition of narrow vibrational resonances and high quantum yield \cite{ferrando-soria_modular_2016}.

On the other side, there is evidence that centrosymmetric molecules spontaneously break the "IR-Raman" selection rule when embedded in a solid-state matrix \cite{myers_vibronic_1994, kulzer_single-molecule_1997}.
Regarding the DBT molecule, a hint comes from the observation that in such conditions they can have a permanent dipole moment, as confirmed by measuring the linear component of the Stark shift \cite{pazzagli_self-assembled_2018, duquennoy_enhanced_2024}, signaling a clear break of centrosymmetry (the amount of permanent dipole moment can be engineered by specifically designing the matrix to maximize the asymmetry \cite{faez_design_2015, moradi_matrixinduced_2019}).
At the same time, other studies have shown that in the matrix the molecule can also exhibit Raman-active modes that are not present in its free space configuration \cite{zirkelbach_high-resolution_2022}. This suggests that the matrix-induced symmetry-breaking may also activate vibrational modes that are typically forbidden.
Even with a strongly reduced Franck-Condon overlap, these "IR-Raman" active vibrational modes could be used to realize our current proposal.
In this perspective, probing the transduction with a broadband quantum cascade laser (QCL) \cite{senica_broadband_2023} also represents a viable way to do THz spectroscopy on the molecule, being able to discern the symmetry or asymmetry of each different vibrational modes.

\subsection{THz electric field amplitude}
As a final step of this experimental characterization, we consider the problem related to the THz electric field intensity $E_{\rm THz}$.
Considering the relation between radiation intensity and the electric field in Eq. \eqref{eq:intensity_Efield}, we estimate that for $I\sim 0.1$kW$/$cm$^2$ of THz radiation, we have an average RMS electric field $E_{\rm THz} \sim 10^3\,$V$/$m. Assuming a vibrational dipole strength $\xi_{\rm v}\sim 10^{-3}\,$nm, we have 
\begin{equation}\label{eq:Rabi_THz_estimation_free_radiation}
\frac{\Omega_{\rm THz}}{2\pi} = \xi_{\rm v} \frac{eE_{\rm THz}}{2\pi\hbar}  \sim 20\,{\rm kHz}.
\end{equation}
From the previous analysis, it is clear that, to have a detectable signal, we require $\Omega_{\rm THz}\gtrsim \gamma_{0}$. Using $\gamma_{0}$ of DBT molecule as a reference, it means that we need $\Omega_{\rm THz}/(2\pi)\sim 10-100\,$MHz. 
The value obtained in Eq. \eqref{eq:Rabi_THz_estimation_free_radiation} for the THz free space radiation is too small.
Assuming $\chi_{\rm trans}\sim 10^{-3}$, and with a detector click probability $p_{\rm click}\sim 0.05$, as in the previous section, this would correspond to a completely negligible transduced-fluorescence rate $p_{\rm click}\Gamma_{\rm trans}/(2\pi)=\chi_{\rm trans}\Omega_{\rm THz}^2/(2\pi\gamma_0)\sim 10^{-7}\,$kcps.
We need then to consider new setups where the RMS THz electric field can be boosted by at least a few orders of magnitude.

\begin{figure}
    \centering
    \includegraphics[width=\columnwidth]{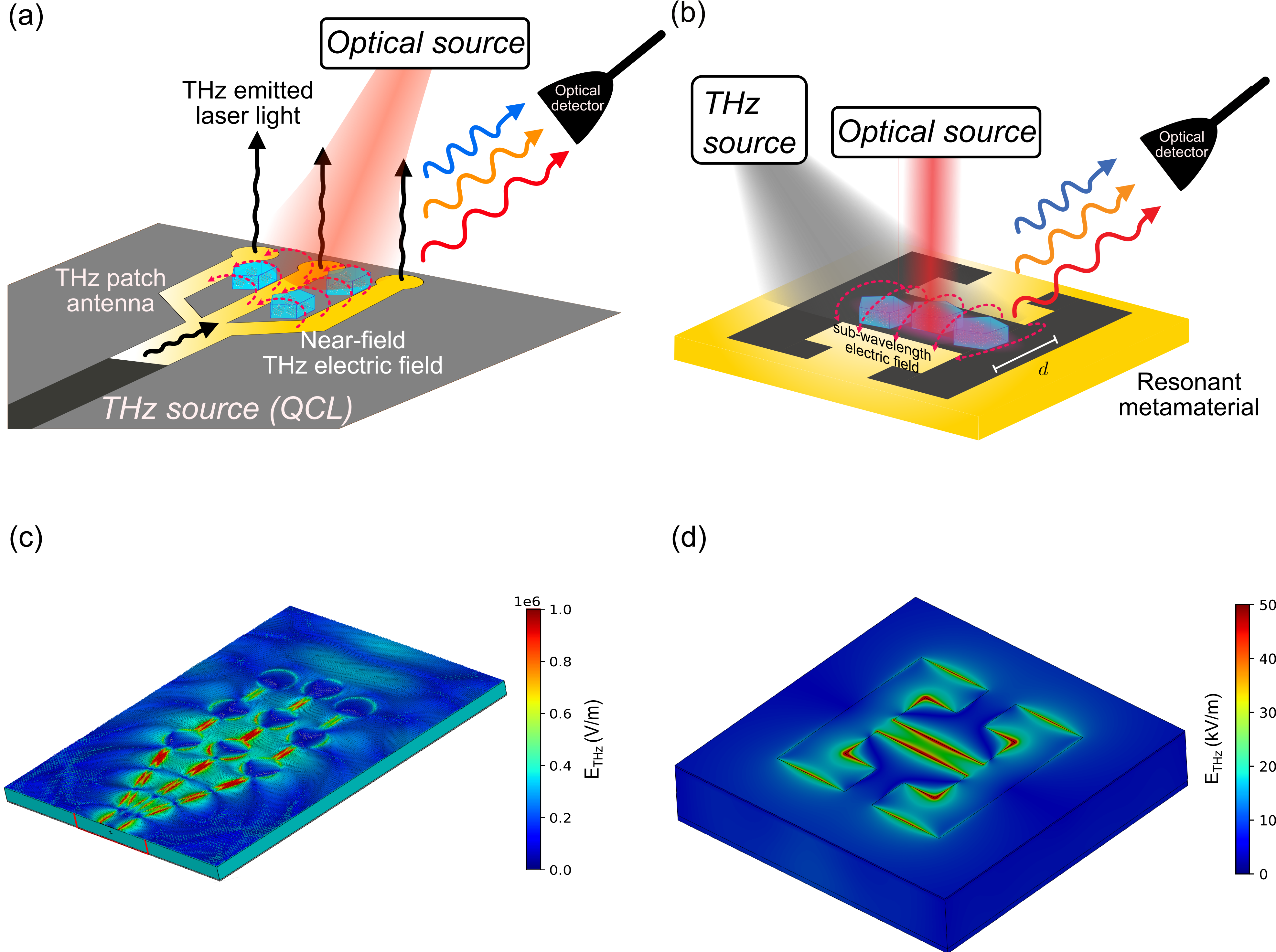}
    \caption{Representation of possible transduction setups. (a) Molecules trapped in matrix nano-crystals are directly placed in contact on the patch antenna of a quantum cascade laser. The molecules are directly coupled to the intense electric near-field. (b) Molecules trapped in matrix nano-crystals placed in the gap of a THz resonant meta-material (in this sketch a split-ring resonator). The external THz radiation is absorbed by the resonant meta-material, and the oscillating THz field is compressed in the sub-wavelength gap, enhancing the electric field strength of a few orders of magnitude.
    (c) Finite element simulation of the electric field amplitude in a patch antenna directly driven by a THz source (QCL), as represented in (a).
    The metal is set as "PEC", and the material is BCB (refractive index 1.57) of height 12 $\mu$m, and PEC also under underneath. The simulation sends the fundamental TM mode as input.
    (d) Finite element simulation of the electric field amplitude in a split-ring resonator driven by a incident plane wave, as represented in (b). The metal is again "PEC" over a substrate of GaAs. The resonator gap $d=4\,\mu$m, and the resonant frequency is $\omega_{\rm LC}/(2\pi)=3.5\,$THz.
The simulations are performed using CST Studio Suite.}
    \label{fig:9}
\end{figure}

The first example is schematically represented in Fig. \ref{fig:9}(a).
Here, the molecules (enclosed in a nano-crystal matrix \cite{pazzagli_self-assembled_2018}) are placed directly on the patch-antenna of a THz radiation source \cite{senica_broadband_2023}.
In this way, the molecules are coupled directly with the sub-wavelength near-field of the structure rather than its radiative part.
The RMS electric field here can reach values around $E_{\rm near}\sim 10^6\,$V$/$m \cite{scalari_thz_2009,senica_broadband_2023}, as also visible from the finite element simulation in Fig. \ref{fig:9}(c).
Assuming $\xi_{\rm v}\sim 10^{-3}\,$nm we have a Rabi frequency 
\begin{equation}\label{eq:THz_Rabi_nearfield}
\frac{\Omega_{\rm THz}^{\rm patch}}{2\pi} =\xi_{\rm v} \frac{eE_{\rm near}}{2\pi\hbar}  \sim 0.2\,{\rm GHz}.  
\end{equation}
The transduced fluoresce signal here is quite large and can easily reach values as $p_{\rm click}\Gamma_{\rm trans}/(2\pi)=p_{\rm click}\chi_{\rm trans}(\Omega_{\rm THz}^{\rm patch})^2/(2\pi\gamma_0)\sim 50\,$kcps.
The possibility to employ on-chip coherent THz sources as quantum cascade laser frequency combs represents a valuable proof of principle with parameters within the optimal range for an eventual molecular THz-optical transducer. It offers the intriguing possibility of studying a fully coherent process \cite{senica_planarized_2022}.
On the other side, this setup faces the restriction that the molecule must be coupled to the near-field, and it is thus unsuited to be used for radiation detection.

This limitation can be overcome by the setup described in Fig. \ref{fig:9}(b).
The THz radiative field is collected by an antenna, which compresses it into a sub-wavelength region, where it can finally couple with the molecule.
This configuration could be naturally implemented using a resonant meta-material, such as a split-ring resonator, where the oscillating electric field is confined in a gap of size $d\sim 1-10\,\mu$m  \cite{maissen_ultrastrong_2014, liu_highly_2015, rajabali_ultrastrongly_2022}.
As described in full detail in App. \ref{app:thz_metamaterial}, the THz Rabi frequency is then boosted by a \emph{capacitive gain} factor that can reach values up to $G_C\approx 100$.
Considering the value of the radiative Rabi frequency estimated in Eq. \eqref{eq:Rabi_THz_estimation_free_radiation}, we then obtain
\begin{equation}\label{eq:Rabi_THz_estimation_free_radiation}
\frac{\Omega_{\rm THz}^{\rm meta}}{2\pi} = G_C \frac{\Omega_{\rm THz}}{2\pi}  \sim 2\,{\rm MHz}.
\end{equation}
This value is still much smaller than coupling the molecule directly with the near-field, as in Eq. \eqref{eq:THz_Rabi_nearfield}, due to the electric field amplitude in the gap of around $E_{\rm gap}\sim 10^4$V/m (as visible from the finite element simulation in Fig. \ref{fig:9}(d)).
Indeed, it produces a signal around $p_{\rm click}\Gamma_{\rm trans}/(2\pi)\sim 0.005\,$kcps within the same assumption for the other estimates given above.
However, we stress that in the absence of THz radiation, no light is emitted at the detected optical frequencies, allowing the background noise from the molecule itself to be, in principle, much smaller than this counting rate.
Leveraging over the collective enhancement, for which one can expect $N_{\rm mol}\sim 10-20$ resonant per nanocrystal-matrix \cite{pazzagli_self-assembled_2018}, and with the proper improvement of the experimental technology, for instance, with a sensible increase of $p_{\rm click}$, this signal can be detected.

Despite the high technological difficulty, it is also worth considering the case when the molecule is directly embedded in a ultra-narrow gap, that can achieve sub-microns distances $d\sim 10-100\,$nm \cite{seo_terahertz_2009, keller_few-electron_2017, lee_more_2023}.
In this case the capacitive gain reaches the maximum value of $G_C\approx10^4$, increasing the THz Rabi frequency up to $\Omega_{\rm THz}^{\rm meta}/(2\pi)\sim 0.2\,$GHz. This value is the same as what was obtained in the near-field coupling in Eq. \eqref{eq:THz_Rabi_nearfield}, but is only due to sub-wavelength compression of the resonant metamaterial.


\section{Conclusions }
\label{sec:conclusion}

In this work, we present a basic theory that shows how to coherently interface vibrational and electronic degrees of freedom in fluorescent organic molecules.

Focusing on a single vibrational mode and combining laser sources resonant with the ZPL, Stokes, and anti-Stokes frequencies, the molecule's dynamics can be manipulated to recover various paradigmatic cavity QED models, such as the Jaynes-Cummings and the quantum Rabi model.
Here, the vibrational mode plays the role of the Bosonic cavity mode.
The competition between the engineered Hamiltonian dynamics and the strong vibrational dissipation can lead to non-trivial steady-states exhibiting vibrational bi-modality, having the feature of a statistical mixture of cat states.

This behavior can be further appreciated by unraveling the system's density matrix into its quantum trajectories, directly observing its composition in vibrational cat states.
Using another supplemental laser, tuned to what we called anti-Stokes$^{\mathbi{2}}$ frequency, we proposed to activate a new hybrid process involving two vibrational quanta that break the symmetry of the quantum Rabi Hamiltonian. 
As a consequence, vibronic quantum jumps can be triggered in the bi-modal state, which are visible as blinking in the anti-Stokes photon counting or in an eventual homodyne detection of the ZPL fluorescence.

By combining the hybrid quantum-state engineering framework with a direct THz driving, one can exploit the Stokes coupling to implement a THz-to-optical transducer.
Again, the theory is developed in full analogy with cavity QED, characterizing the transducer functionality by introducing the vibrational cooperativity and saturation number.
After expressing its efficiency through a linear response function, we assess that molecules with high quantum yield and narrow vibrational transitions turn out to be optimal candidates as THZ-to-optical transducers.

Fluorescent PAH molecules would be good options, although facing the fundamental problem of being mostly centro-symmetric. 
This symmetry forbids the molecule to have THz absorption simultaneous to the Stokes Raman transition, effectively forbidding the transduction.
However, as for spin-active molecules \cite{ferrando-soria_modular_2016,gaita-arino_molecular_2019, wasielewski_exploiting_2020}, there might be an interest for the community to engineer and build PAH-like molecules with the desired asymmetry with the goal of improving the THz sensitivity.
Another possibility is to explore, and eventually exploit, the asymmetry induced by the molecule's insertion into the solid-state matrix.

Because of the extremely long wavelength of THz radiation, the expected possible molecule's cross-section is extremely small.
To circumvent this limitation one can place the molecule directly on the surface of the THz source and then couple it with the intense near-field of the output antenna.
Another strategy may instead involve the use of a metamaterial resonant structure acting as a buffer.
A THz resonator (like a split-ring cavity) can match the impedance between free space THz radiation and the extremely sub-wavelength molecular vibrational dipole transition sensibly boosting its absorption.

Overall, our analysis shows that the combination of electronic and vibrational transitions in fluorescent molecules is an extremely valuable tool for quantum science and technology \cite{gurlek_small_2024}.
It can be used to explore non-classical states of mechanical degrees of freedom in strict analogy with the field of optomechanics \cite{aspelmeyer_cavity_2014, barzanjeh_optomechanics_2022}, but it may also be an important piece for THz technology at the single quantum level \cite{todorov_thz_2024}.



\acknowledgements{The project has been co-funded by the European Union (ERC, QUINTESSEnCE, 101088394). Views and opinions expressed are however those of the author(s) only and do not necessarily reflect those of the European Union or the European Research Council. Neither the European Union nor the granting authority can be held responsible for them. 
We are grateful to Claudiu Genes, Diego Martin-Cano, Quentin Deplano,  Francesco Campaioli, Alberto Biella, Fabrizio Minganti and Severino Zeni for very insightful discussions. 
D.D.B. acknowledges funding from the European Union - NextGeneration EU, "Integrated infrastructure initiative in Photonic and Quantum Sciences" - I-PHOQS [IR0000016, ID D2B8D520, CUP B53C22001750006].
}

\appendix

\section{Frequency filtered photodetection}
\label{app:frequency_filter}

Equivalently to the theory described in Ref. \cite{del_valle_theory_2012}, we can describe the frequency-resolved photodetection within the master equation approach by including a bath with structured density of states $\mathcal{F}(\omega)$, peaked at the frequencies of interest, and normalized such that $\mathcal{F}(\omega)\leq  1$.
We refer to $\mathcal{F}(\omega)$ also as the detector's filter spectral density.
In this description, the detector is the bath itself.

To include the frequency resolution, we take the approach described in Ref. \cite{beaudoin_dissipation_2011, di_stefano_photodetection_2018}.
Assuming that the system is coupled to the detector through the system's generic operator $\hat{S}$.
As explained in Ref. \cite{di_stefano_photodetection_2018}, we need to split this operator into its positive and negative frequency components by introducing
\begin{equation}
    \hat{S}_{nm} = \sum_{u,v\in \mathbb{D}(\omega_{mn})}|u\rangle \langle{u | \hat{S} | v \rangle} \langle v |.
\end{equation}
This is the spectrally filtered jump operator between the system's energy eigenstates, where $|n\rangle, |m\rangle$ is the $n,m$-th eigenstate with $\omega_n, \omega_m$ its corresponding eigenfrequency of the total system's Hamiltonian, and $\omega_{mn}=\omega_m-\omega_n$ are the system's Bohr frequencies.
Notice that here we have summed over the set $\mathbb{D}(\omega_{mn})$ of possible degenerate transition-states $|u\rangle,|v\rangle$ such that $|\omega_{vu}-\omega_{mn}|<\kappa_{\rm D}$, as customary to treat degenerate transitions in the master equation derivation \cite{cattaneo_local_2019}.
Since $\sum_{n<m}(\hat{S}_{nm} + \hat{S}_{nm}^{\dag}) = \hat{S}$, this is just a way to split each jump operator in its \emph{positive frequency} components \cite{cohentannoudji_atomphoton_1998}. 
As pointed out in Ref. \cite{beaudoin_dissipation_2011}, to derive Eq. \eqref{eq:det_linbladian}, one assumes an independent, individual bath for each transition $n,m$. Each of these baths has flat density of states and a bandwidth given by $\kappa_{\rm D}$, essentially being a detector only for the $n,m$-transition, thus constituting a frequency resolution pixel of the whole detector.
In this perspective, the bandwidth of the detector is set by the region in $\omega$ where $\mathcal{F}(\omega)\approx 1$, while the minimum detector resolution is set by $\kappa_{\rm D}$.

Since every $n,m$-transition is independent by all the non-degenerate others, one can use the input/output relations \cite{gardiner_physics_2015} to express all the observables seen by the detector as products and sum of the spectrally filtered jump operators weighted by the bath density of states $\sqrt{\mathcal{F}(\omega_{mn})}\hat{S}_{nm}$.
For instance, the total energy absorbed by the detector is 
\begin{equation}
    W = \frac{\hbar \kappa_{\rm D}}{4}\sum_{\vec{\sigma}}\mathcal{F}(\omega_{\vec{\sigma}})\langle{ \hat{S}_{\vec{\sigma}}^{\dag}\hat{S}_{\vec{\sigma}}  \rangle},
\end{equation}
from which we define the photon number operator
\begin{equation}
    \hat{N}_{\mathcal{F}} = \sum_{\vec{\sigma}}\mathcal{F}(\omega_{\vec{\sigma}})\langle{ \hat{S}_{\vec{\sigma}}^{\dag}\hat{S}_{\vec{\sigma}}  \rangle}.
\end{equation}

The frequency-filtered emission spectrum (resonance fluorescence) is given by
\begin{equation}\label{eq:S_filter_general}
    \mathcal{S}_{\rm filter}(\omega ) = \sum_{\vec{\sigma}} \mathcal{F}(\omega_{\vec{\sigma}}) \int dt e^{i\omega t} \langle{\hat{S}_{\vec{\sigma}}^{\dag}(t) \hat{S}_{\vec{\sigma}} \rangle},
\end{equation}
and the photon coincidences 
\begin{equation}\label{eq:g2_filter_general}
    g^{(2)}_{\rm filter}(\tau) = \sum_{\vec{\sigma},\vec{\lambda} } \mathcal{F}(\omega_{\vec{\sigma}})\mathcal{F}(\omega_{\vec{\lambda}}) \frac{\langle{\hat{S}_{\vec{\sigma}}^{\dag} \left[ \hat{S}_{\vec{\lambda}}^{\dag}\hat{S}_{\vec{\lambda}}\right](\tau) \hat{S}_{\vec{\sigma}}  \rangle}}{ \langle{\hat{S}_{\vec{\sigma}}^{\dag}\hat{S}_{\vec{\sigma}}\rangle} \langle{\hat{S}_{\vec{\lambda}}^{\dag}\hat{S}_{\vec{\lambda}} (\tau)\rangle} }. 
\end{equation}
Here $\vec{\sigma}=(n,m), \vec{\lambda}=(p,q)$ with $n<m$ and $p<q$, are vectorial indices to shorten the notation.

The back-action on the system due to the detection events is then given by the detector Lindbladian dissipator
\begin{equation}\label{eq:det_linbladian}
    \mathcal{L}_{\rm det}(\hat{\rho}) = \frac{\hbar \kappa_{\rm D}}{2}\sum_{n<m}\mathcal{F}(\omega_{mn})\left[ 2 \hat{S}_{nm}\, \hat{\rho}\, \hat{S}_{nm}^{\dag} - \lbrace{\hat{S}_{nm}^{\dag}\hat{S}_{nm},\, \hat{\rho} \rbrace}  \right],
\end{equation}
It is worth noticing that if $\kappa_{\rm D} \ll \gamma_{0}\ll \gamma_{\rm v}$, the dissipative back-action of the detector on the system is completely negligible. In this respect, we can thus neglect the contribution of Eq. \eqref{eq:det_linbladian} to the whole molecule's master equation and just access the detected quantities by the above-mentioned expectation values.

The frequency component decomposition of the jump operators is a treatment typically used in the field of ultra-strong coupling and non-perturbative cavity QED; see, for instance, Ref. \cite{beaudoin_dissipation_2011,de_bernardis_tutorial_2024}.
Most often, splitting the jump operators in this way cannot be represented analytically, and it requires to numerically diagonalize the Hamiltonian and numerically reconstruct the positive frequency components of the jump operator.
Fortunately, thanks to the polaron transformation, here we can represent everything analytically.
After the polaron transformation in Eq. \eqref{eq:polaron_mapping_operators}, and taking $\hat{S}=\hat{\sigma}_{\rm pol}$ as the jump operator coupled to the detector (and the electromagnetic environment as well), it is straightforward to split it when represented on the eigenbasis of the polaron Hamiltonian in Eq. \eqref{eq:Ham_mol_polaron}. 
Using the Taylor series of the displacement operators in Eq. \eqref{eq:displacement_normal_order_expansion} and assuming $\omega_{\rm v}\ll \omega_0$ together with $\eta < 1$, the polaron jump operator $\hat{S}=\hat{\sigma}_{\rm pol}$ is immediately expressed as a sum of positive frequency terms
\begin{equation}\label{eq:polaron_jump_series_exp}
\begin{split}
    \hat{\sigma}_{\rm pol}(t) & = e^{-\frac{\eta^2}{2}}\sum_{n,m} \frac{(\eta \hat{b}^{\dag})^n}{n!} \frac{(-\eta \hat{b})^m}{m!} \hat{\sigma} e^{-i(\omega_0 - (n-m)\omega_{\rm v})t} 
    \\
    &= \sum_{n,m} \hat{S}_{nm}(t).
\end{split}
\end{equation}
Here, the polaron lowering operator is taken in interaction picture with respect to the polaron system's Hamiltonian in Eq. \eqref{eq:Ham_mol_polaron} highlighting the split in different frequency components. We clearly see that $\omega_0 + (n-m)\omega_{\rm v} > 0$ for each $n,m\in \mathbb{N}$ provided that $m-n < \omega_0/\omega_{\rm v}\sim 70$. The last estimate is based on the values of $\omega_0$ and $\omega_{\rm v}$ for a typical DBT molecule \cite{zirkelbach_high-resolution_2022} and is somehow similar for many fluorescent molecules.
Having $\eta \ll 1$ makes the contributions at negative frequencies completely negligible, so we can take \eqref{eq:polaron_jump_series_exp} as the decomposition of the polaron lowering operator in its positive frequency components.

In case of strong driving, for instance, including in the Hamiltonian the driving terms in Eqs. \eqref{eq:ham_drive_zpl}-\eqref{eq:ham_drive_stokes}-\eqref{eq:ham_drive_antistokes}, when the driving Rabi frequencies are larger than the detector bandwidth $\Omega>\kappa_{\rm D}$, one needs to re-diagonalize the Hamiltonian including these driving terms. The jump operator must be re-written on this new eigenbasis, including the driving dressing.
However, this leads to observable consequences only if one can resolve the difference of the new positive frequency terms, requiring a detector with resolution at the scale of the driving Rabi frequencies $\Omega$ \cite{wiseman_are_2012}. 
For our specific purpose, we do not account for this possibility, which is instead left for a future work.

Taking only the linear order in $\eta$, we reduce to the case discussed in the main text.
For a flat filter function $\mathcal{F}(\omega) = 1$ (no filtering) and assuming the simple case where optical and vibrational excitations are uncorrelated and factorized, the emission spectrum becomes
\begin{equation}\label{eq:emission_spectrum_analitical}
\begin{split}
    \mathcal{S}(\omega) \approx \mathcal{S}_{\rm zpl}(\omega) & + \eta^2  \int d\omega'  \mathcal{S}_{\rm S}(\omega-\omega') \mathcal{S}_{\rm zpl}(\omega') 
    \\
    &+  \eta^2 \int d\omega'  \mathcal{S}_{\rm AS}(\omega-\omega') \mathcal{S}_{\rm zpl}(\omega'),
\end{split}
\end{equation}
where 
\begin{equation}
    \mathcal{S}_{\rm zpl}(\omega) = \int dt e^{i\omega t}\langle{\hat{\sigma}^{\dag}(t) \hat{\sigma}\rangle},
\end{equation}
\begin{equation}
    \mathcal{S}_{\rm S}(\omega) = \int dt e^{i\omega t}\langle{\hat{b}(t) \hat{b}^{\dag}\rangle},
\end{equation}
\begin{equation}
    \mathcal{S}_{\rm AS}(\omega) = \int dt e^{i\omega t}\langle{\hat{b}^{\dag}(t) \hat{b}\rangle}.
\end{equation}
In the simplest case of uncoupled vibrational and electronic degrees of freedom, we have $\hat{b}(t) = \hat{b}e^{-i\omega_{\rm v} t}$.
We thus have that $\mathcal{S}_{\rm S}(\omega) = \delta(\omega+\omega_{\rm v}) + \mathcal{S}_{\rm AS}(-\omega)$.
Assuming a steady-state Lorentzian profile for both ZPL and anti-Stokes, $\mathcal{S}_{\rm zpl}(\omega)={\rm L}(\omega-\omega_0, \gamma_{0})$, $\mathcal{S}_{\rm AS}(\omega)={\rm L}(\omega-\omega_{\rm v}, \gamma_{\rm v})$, we have
\begin{equation}
\begin{split}\label{eq:exp_S}
    \mathcal{S}(\omega) \approx &  {\rm L}(\omega-\omega_0, \gamma_{0}) + \eta^2 n_{\rm v} {\rm L}(\omega-(\omega_0+\omega_{\rm v}), \gamma_{\rm v} + \gamma_{0})
    \\
    & + \eta^2(1+n_{\rm v}){\rm L}(\omega-(\omega_0-\omega_{\rm v}), \gamma_{\rm v} + \gamma_{0}).
\end{split}
\end{equation}
Here ${\rm L}(\omega-\omega_{\rm c}, \gamma)$ is a normalized Lorentzian profile centered in $\omega_{\rm c}$ and with full-width-half-maximum (FWHM) $\gamma$, and $n_{\rm v} = \braket{\hat{b}^{\dag}\hat{b}}$ is the vibrational steady-state occupation.

Experimentally one can extract the $\eta$ parameter by measuring the area underlying the ZPL and Stokes peaks in Eq. \eqref{eq:emission_spectrum_analitical} (assuming the profile is not normalized), $A_{\rm zpl}= \int d\omega \mathcal{S}_{\rm zpl}(\omega)$, $A_{\rm S} =\int d\omega \mathcal{S}_{\rm S}(\omega)$, $\eta^2 \approx A_{\rm S}/A_{\rm zpl}$ (the result is exact only if the profiles are Lorentzian as in Eq. \eqref{eq:exp_S}).
We call this quantity Franck-Condon factor.

Notice that sometimes the Franck-Condon factor referes to the ratio between the Stokes area and the whole integrated spectrum, $A_{\rm S}/(A_{\rm zpl}+A_{\rm S})\approx \eta^2/(1+\eta^2)$.

\section{Master equation}
\label{app:master}

As described in Ref. \cite{clear_phonon-induced_2020}, the driven-dissipative system is well described by the  master equation
\begin{equation}\label{eq:master_equation_general_0}
    \hbar \partial_t \hat{\rho} = \mathcal{L}_{H}(\hat{\rho} ) + \mathcal{L}_{\gamma_{0}}(\hat{\rho}) + \mathcal{L}_{\gamma_{\rm v}}(\hat{\rho}) + \mathcal{L}_{\phi_{0}}(\hat{\rho})  + \mathcal{L}_{\phi_{\rm v}}(\hat{\rho}).
\end{equation}
In the right-hand-side, each term represents respectively: the Hamiltonian dynamics, the electronic spontaneous decay, the vibrational spontaneous decay, the electronic dephasing, and the vibrational dephasing.

After the polaron transformation, the coherent dynamics (including the drivings) is given by all the Hamiltonian terms
\begin{equation}
    \mathcal{L}_{H}(\hat{\rho} ) = -i\left[ H_{\rm mol} + H_{\rm opt} + H_{\rm THz}, \hat{\rho}  \right].
\end{equation}
When the Franck-Condon factor is zero $\eta=0$, the electronic and vibrational dissipation are simply given by
\begin{equation}
    \mathcal{L}_{\gamma_{0}}(\hat{\rho}) = \frac{\hbar \gamma_{0}}{2}\left[ 2\hat{\sigma}\, \hat{\rho} \, \hat{\sigma}^{\dag} - \lbrace{ \hat{\sigma}^{\dag}\hat{\sigma},\, \hat{\rho} \rbrace} \right],
\end{equation}
\begin{equation}
    \mathcal{L}_{\gamma_{\rm v}}(\hat{\rho}) = \frac{\hbar \gamma_{\rm v}}{2}\left[ 2\hat{b}\, \hat{\rho} \, \hat{b}^{\dag} - \lbrace{ \hat{b}^{\dag}\hat{b}, \hat{\rho} \rbrace} \right].
\end{equation}
However, for non-zero Franck-Condon factor, $\eta\neq 0$, the electronic and vibrational degrees of freedom are coupled, and these two terms need to be re-derived.
With the help of the polaron transformation, and by introducing the electronic bath density of states $J_0(\omega)$, this can be immediately done by the substitution
\begin{equation}
    \hat{\sigma} \longmapsto e^{-\frac{\eta^2}{2}}\sum_{n,m} \sqrt{J^{nm}_0}\frac{(\eta \hat{b}^{\dag})^n}{n!} \frac{(-\eta \hat{b})^m}{m!} \hat{\sigma},
\end{equation}
where, for brevity, we define $J^{nm}_0 = J_0(\omega_0 - (n-m)\omega_{\rm v})$.
The same is repeated for the vibration
\begin{equation}
    \hat{b} \longmapsto \sqrt{J_{\rm v}(\omega_{\rm v})}\hat{b} + \eta \sqrt{J_{\rm v}(0)}\hat{\sigma}^{\dag}\hat{\sigma},
\end{equation}
where $J_{\rm v}(\omega)$ is the vibrational bath density of states. Since at low frequency the bath density of states is always assumed to go to zero $J_{\rm v}(\omega=0)=0$, we can safely neglect the last term of the equation above.
The polaron transformed electronic and vibrational decay Lindbladian are then given by
\begin{equation}
\begin{split}
    \mathcal{L}_{\gamma_{0}}(\hat{\rho}) &= \frac{\hbar \gamma_{0}}{2} e^{-\eta^2} 
    \\
    &\times \sum_{n,m} J^{nm}_0 \frac{\eta^{2n}(-\eta)^{2m}}{(n!m!)^2} \left[ 2\hat{p}_{nm}\, \hat{\rho} \, \hat{p}_{nm}^{\dag} - \lbrace{ \hat{p}_{nm}^{\dag}\hat{p}_{nm},\, \hat{\rho} \rbrace} \right]
    \\
    &\approx J_0(\omega_0)\frac{\hbar \gamma_{0}}{2} \left[ 2\hat{\sigma}\, \hat{\rho} \, \hat{\sigma}^{\dag} - \lbrace{ \hat{\sigma}^{\dag}\hat{\sigma},\, \hat{\rho} \rbrace} \right] + O(\eta^2)
\end{split}
\end{equation}
\begin{equation}
    \mathcal{L}_{\gamma_{\rm v}}(\hat{\rho}) \approx J_{\rm v}(\omega_{\rm v})\frac{\hbar \gamma_{\rm v}}{2}\left[ 2\hat{b}\, \hat{\rho} \, \hat{b}^{\dag} - \lbrace{ \hat{b}^{\dag}\hat{b}, \hat{\rho} \rbrace} \right].
\end{equation}
Here $\hat{p}_{nm} = \left(\hat{b}^{\dag}\right)^n\hat{b}^m \hat{\sigma}$ is a single-frequency polaron component.
The last approximation in the electronic Lindbladian is well justified when $\gamma_0$ is the smallest frequency rate in the system. For a typical Franck-Condon factor $\eta^2 \ll 1$, the mixed vibrational/electronic decays due to the higher terms in the polaron expansion are then completely negligible.
Having the two densities of states normalized such that $J_0(\omega_0)=J_{\rm v}(\omega_{\rm v})=1$, one recovers the result in the main text.

Focusing our interest on molecules trapped in solid-state matrices, it would be also important to consider the photon and vibration dephasing terms, given by
\begin{equation}
    \mathcal{L}_{\phi_{\rm opt}}(\hat{\rho}) = \frac{\hbar \gamma_{\phi, \rm opt}}{2}\left[ 2 \hat{\sigma}^{\dag}\hat{\sigma}\, \hat{\rho}\, \hat{\sigma}^{\dag}\hat{\sigma} - \lbrace{ \hat{\sigma}^{\dag}\hat{\sigma},\,  \hat{\rho} \rbrace} \right],
\end{equation}
\begin{equation}
    \mathcal{L}_{\phi_{\rm v}}(\hat{\rho}) = \frac{\hbar \gamma_{\phi, \rm v}}{2}\left[ 2 \hat{b}^{\dag}\hat{b}\, \hat{\rho}\, \hat{b}^{\dag}\hat{b} - \lbrace{ \hat{b}^{\dag}\hat{b},\,  \hat{\rho} \rbrace} \right].
\end{equation}
Here $\gamma_{\phi, \rm opt}=\gamma_{\phi, \rm opt}(T)$, $\gamma_{\phi, \rm v}=\gamma_{\phi, \rm v}(T)$ are the temperature dependent optical and vibrational dephasing rates.
When the system operates at cryogenic temperatures, around $T\sim 1$K, these dephasing rates are exponentially suppressed \cite{clear_phonon-induced_2020}, so unless specified, we always neglect the effect of dephasing.

\section{Saturation curves}
\label{app:sat_curve}

The emitted fluorescence depends on the scheme adopted to excite the molecule and the scheme to detect the emitted radiation.
Two among the most popular schemes in the field are resonant excitation of the HOMO-LUMO transition and detection of the Stokes shifted emission, off-resonant excitation via an anti-Stokes Raman process and detection through the zero-phonon line (after the relaxation of the Raman-excited vibrational quantum). 

The first scheme results in a fluorescence rate
\begin{equation}\label{eq:photo_rate_Stokes_zplexc}
    \Gamma_{\rm res} = p_{\rm click} \gamma_{0}\langle{\hat{\mathrm{N}}_{\rm S}\rangle} = p_{\rm click} \gamma_{0} \eta^2 \langle{\hat{\sigma}^{\dag}\hat{\sigma}\rangle}_{\rm res},
\end{equation}
where $p_{\rm click}\in [0;1]$ is proportional to the detector coupling rate $\kappa_D$, accounting for the probability that the detector is activated by the photon.
The excited population is derived by solving the related optical-Bloch equations related to the ZPL driving Hamiltonian in Eq. \eqref{eq:ham_drive_zpl} inside the general master equation in Eq. \eqref{eq:master_equation_general} \cite{cohentannoudji_atomphoton_1998, haroche_exploring_2006}.
Solving for the steady-state, one obtains
\begin{equation}
    \langle{\hat{\sigma}^{\dag}\hat{\sigma}\rangle}_{\rm res} = \frac{1}{2}\frac{1}{1 + n_{\rm zpl}}.
\end{equation}
Here $n_{\rm zpl} = (4\Delta_0^2 + \gamma_{0}^2)/(2\Omega_{\rm zpl}^2)$ is the ZPL saturation number.

For the incoherent driving scheme, instead, we have 
\begin{equation}\label{eq:photo_rate_incoh}
    \Gamma_{\rm incoh} = p_{\rm click} \gamma_{0}\langle{\hat{\mathrm{N}}_{\rm zpl}\rangle} = p_{\rm click} \gamma_{0} \langle{\hat{\sigma}^{\dag}\hat{\sigma}\rangle}_{\rm incoh},
\end{equation}
missing the Franck-Condon suppression factor $\sim \eta^2$ with respect to the resonant scheme because we are directly detecting the ZPL emission.
As discussed in the main text, we can compute the excited population by tracing away the vibrational mode from the anti-Stokes Hamiltonian \cite{gardiner_physics_2015}, obtaining a new Lindbladian contribution
\begin{equation}\label{eq:Gamma_plus_up_rate}
    \mathcal{L}_{\rm pump}(\hat{\rho}) = \frac{\Gamma_+}{2}\left( 2\hat{\sigma}^{\dag} \hat{\rho} \hat{\sigma} - \lbrace{ \hat{\sigma} \hat{\sigma}^{\dag}, \hat{\rho} \rbrace}\right).
\end{equation}
Deriving its related optical Bloch equations by including it in the total master equation (this time without ZPL driving term), we obtain
\begin{equation}
    \langle{\hat{\sigma}^{\dag}\hat{\sigma}\rangle}_{\rm incoh} = \frac{\Gamma_{+}}{\Gamma_{+}+\gamma_{0}}.
\end{equation}
Here, the pumping rate is given by
\begin{equation}\label{eq:rho_ee_incoherent_drive_anal}
    \Gamma_+ = \frac{g_{\rm AS}^2}{\gamma_{\rm v}} \frac{\gamma_{\rm v}^2}{4\Delta_{0}^2 + \gamma_{\rm v}^2} .
\end{equation}
Since $g_{\rm AS} = \eta \Omega_{\rm AS}$, $\eta^2 \sim 10^{-1}$ and $\gamma_{0}/\gamma_{\rm v}\sim 10^{-3}$, accordingly to App. \ref{app:dimensional_param}, we see that the power required to saturate with the incoherent driving is $\sim 10^4$ times the saturation power in the resonant excitation scheme.

\section{Power intensity, Rabi frequencies and dimensional parameters}
\label{app:dimensional_param}
The electric field of a laser beam can be related to its radiation intensity via the formula \cite{jackson_classical_2013, novotny_principles_2012}
\begin{equation}\label{eq:intensity_Efield}
    I = \frac{c\epsilon_0}{2}E^2,
\end{equation}
where $E$ is intended as the RMS value of the main polarization.

The Rabi frequency resulting by irradiating a molecule with dipole transition $\mu=e\xi$ ($e$ is the elementary charge) is then
\begin{equation}\label{eq:Rabi_power_realtion_general}
    \Omega = \xi \sqrt{\frac{8\pi\alpha_{\rm fs} I}{\hbar}},
\end{equation}
where $\alpha_{\rm fs} = e^2/(4\pi\epsilon_0 \hbar c)= 1/137$ is the fine-structure constant.

Taking inspiration from recent experiments with DBT molecules in Anthracene \cite{toninelli_single_2021}, for the optical drive, we take as a reference intensity
\begin{equation}
    I_s = 30\, {\rm W/cm^2},
\end{equation}
which is the resonant saturation intensity for the DBT HOMO-LUMO transition at the resonant frequency of $\omega_0/(2\pi) \approx 381.9$THz ($\lambda_0 = 785$nm).
The Rabi frequency can be then rewritten as $\Omega = \Omega^s \sqrt{I/I_s}$.
Considering that its transition dipole moment is $\mu_0 \approx 11$D we have that $\xi_0\approx0.23$nm and we thus obtain
\begin{equation}
    \frac{\Omega_{0}^s}{2\pi} \approx 0.028\,{\rm GHz}. 
\end{equation}
Notice that it is useful to keep in mind the unit-converted intensity $I_s/\hbar = 3 {\rm THz^2/\mu m^2}$. 
Its conversion to the estimated output power by considering the beam focused at the diffraction limit is $P_{\rm out} \sim \pi \lambda^2 I$, obtaining $P_{0}^s \sim \pi \lambda_0^2 I_s \approx 1$nW. 
In these type of experiments, the output power of a laser can arrive to tens of mW, so the radiation intensity can arrive up to $I < 10^{7}\times I_s=300\,$MW/cm$^2$, and the relative optical Rabi frequency $\Omega_0/(2\pi) < 88\,$GHz.

The same formulas hold for the THz Rabi frequency, with the only replacement of the transition dipole moment, which has to be derived from the specific vibrational transition. 
The vibrational dipole for fluorescent organic molecules can be estimated to be in the range $\xi_{\rm v} \sim [10^{-3};10^{-1}] \times \xi_0$, giving a much smaller Rabi frequency. Moreover, we must consider that THz radiation is on a much lower intensity scale due to its large wavelength \cite{scalari_thz_2009}. 
Considering a high output power of $P_{\rm THz}\sim 1$mW and $\lambda_{\rm THz}\sim 60\, \mu$m ($\omega_{\rm THz}/(2\pi)=5\,$THz) we have $I_{\rm THz} \sim 10\,$W/cm$^2 = I_s/3$ and a Rabi frequency in the range $\Omega_{\rm THz}/(2\pi) \sim 0.1-10 \,$MHz. 
This value is too small if compared to the typical dissipation of the vibrational modes $\gamma_{\rm v}/(2\pi) \sim 10\,$GHz \cite{clear_phonon-induced_2020, zirkelbach_high-resolution_2022}.

\section{Generalization to N-molecules and M-modes of vibration}
\label{app:generalization_multiple}
The model developed in the previous subsections describes only a single molecule with a single vibrational mode. 
To generalise it to N-many identical molecules and M-many modes, we introduce the index $n=1,2, \ldots N$ to label each molecule and, for each molecule, a set of indices $k_n = 1,2, \ldots M$ to label each local vibrational mode. In this way, we have
\begin{align}
    \hat{\sigma} \mapsto \hat{\sigma}_n && \hat{b} \mapsto \hat{b}_{n, k_n}.
\end{align}
The polaron transformed operators are immediately generalised in the same way, provided that a specific Franck-Condon factor and vibrational frequency for each mode are introduced
\begin{equation}
    \eta, \omega_{\rm v} \longmapsto \eta_{k_n}, \omega_{{\rm v}, k_n}
\end{equation}
The total Liouvillian defined in Eq. \eqref{eq:master_equation_general} can be just substituted with
\begin{equation}
    \mathcal{L}_{\rm tot}(\hat{\rho} ) \longmapsto \sum_{n, k_n} \mathcal{L}_{\rm tot}^{n,k_n}(\hat{\rho} ),
\end{equation}

In this generalization, we completely neglect the interaction between molecules, assuming them as independent objects. 
It is still worth noticing that for a dense ensemble, the polaron dressing also affects the molecule-molecule interaction with contributions at most $\sim \eta_{k_n}^2$, which can be small but non-negligible. This contribution leads to a vibrational interaction between the molecules

Even though all the molecules are identical, they are sensibly distorted by the embedding in the matrix, giving rise to non-negligible inhomogeneous broadening \cite{pazzagli_self-assembled_2018}. As a consequence, we have also indexed all the parameters, which are, in principle, all different for each molecule and each mode.
On the other side, for most of the cases discussed here, we can safely consider the molecules all identical with identical parameters.

All the observables, such as the frequency-filtered photon number in App. \ref{app:frequency_filter} are generalized in the same way $\hat{\mathrm{N}}\mapsto \hat{\mathrm{N}}_{n, k_n}$.

For instance, the multi-mode emission spectrum of a single molecule represented in Fig. \ref{fig:1} is given by by
\begin{equation}\label{eq:multi_mode_S}
\begin{split}
        \mathcal{S}(\omega ) = &  {\rm L}(\omega-\omega_0, \gamma_{0}) + \sum_k \eta_k^2 n_k {\rm L}(\omega-(\omega_0+\omega_{k}), \gamma_{k} + \gamma_{0})
    \\
    & + \sum_k \eta_k^2(1+n_k){\rm L}(\omega-(\omega_0-\omega_{k}), \gamma_{k} + \gamma_{0}). 
\end{split}
\end{equation}

\section{Mean-field cavity QED equations}
\label{app:mean_field_OBE}

Here, we derive the semi-classical steady state of the following Hamiltonian
\begin{equation}
\begin{split}
    H \approx & \hbar \Delta_0 \hat{\sigma}^{\dag}\hat{\sigma} + \hbar \Delta_{\rm v} \hat{b}^{\dag}\hat{b} + \frac{\hbar g_{\rm S}}{2}\left(\hat{\sigma} \hat{b}^{\dag} + {\rm h.c.} \right)
    \\
    & + \frac{\hbar \Omega_{\rm zpl}}{2}\left(\hat{\sigma} + {\rm h.c.} \right) + \frac{\hbar \Omega_{\rm THz}}{2}\left( \hat{b}e^{i\theta} + {\rm h.c.} \right),
\end{split}
\end{equation}
where $\Delta_0 = \omega_0 - (\omega_{\rm S}+\omega_{\rm THz})$, $\Delta_{\rm v} = \omega_{\rm v}-\omega_{\rm THz}$ and we assume $\omega_{\rm zpl} \approx \omega_{\rm S}+\omega_{\rm THz}$.

From the full master equation in Sec. \ref{sec:master_eq}, discarding the dephasing terms, we derive the semiclassical mean-field equations
\begin{equation}
    \begin{split}
        i\partial_t b & = \left( \Delta_{\rm v} - i\frac{\gamma_{\rm v}}{2} \right)b + \frac{g_{\rm S}}{2}\sigma + \frac{\Omega_{\rm THz}}{2}
        \\
        i\partial_t \sigma & = \left(\Delta_0 - i\frac{\gamma_0}{2}\right)\sigma - g_{\rm S}s_z \left(b + b_{\rm zpl} \right)
        \\
        \partial_t s_z & = i\frac{g_{\rm S}}{2}\left[\sigma (b+b_{\rm zpl})^* - {\rm h.c.} \right] - \frac{\gamma_{0}}{2} -\gamma_{0}s_z.
    \end{split}
\end{equation}
Here $b=\langle{\hat{b}\rangle}$, $\sigma=\langle{\hat{\sigma}\rangle}$, $s_z=\langle{\hat{\sigma}^{\dag}\hat{\sigma}\rangle}-1/2$, and, without loss of generality, we set $\theta=0$.
Moreover, we re-expressed the ZPL driving term as an effective vibrational shift 
\begin{equation}
    b_{\rm zpl} = \frac{\Omega_{\rm zpl}}{g_{\rm S}}.
\end{equation}

Using the definitions in Eqs. \eqref{eq:coop_satnum_stokes}, the steady state of these equations is given by the following algebraic equations
\begin{equation}\label{eq:bistab_b}
    \left[ 1 + 2i\frac{\Delta_{\rm v}}{\gamma_{\rm v}}  + \frac{\mathcal{C}_{\rm S}}{1+2i\Delta_{0}/\gamma_{0}} \frac{ 1 }{1+|b_{ss}|^2/n_{\rm S}}\right]b_{ss} = b_{\rm in},
\end{equation}
\begin{equation}\label{eq:bistab_rhoee}
   \braket{\hat{\sigma}^{\dag}\hat{\sigma}}_{ss} = \frac{1}{2}\frac{|b_{ss}|^2/n_{\rm S}}{1+|b_{ss}|^2/n_{\rm S}}.
\end{equation}
Here $b_{ss}=\langle{\hat{b}\rangle} + b_{\rm zpl}$, $\braket{\hat{\sigma}^{\dag}\hat{\sigma}}_{ss}$ are the steady-state expectation values, and 
\begin{equation}
    b_{\rm in} = -i\frac{\Omega_{\rm THz}}{\gamma_{\rm v}} + \left[ 1 + 2i\frac{\Delta_{\rm v}}{\gamma_{\rm v}}\right]b_{\rm zpl}
\end{equation}
is the total input driving amplitude accounting for both the THz drive and the ZPL drive.
These expressions generalize the saturation formulas described in App. \ref{app:sat_curve}.

At full resonance $\Delta_{0}=\Delta_{\rm v}=0$, and assuming small vibrational amplitude $b_{ss}\ll n_{\rm S}$, we expand Eq. \eqref{eq:bistab_b}, obtaining
\begin{equation}\label{eq:general_b_avg}
    \langle{\hat{b}\rangle} \approx \frac{\mathcal{C}_{\rm S} }{1+\mathcal{C}_{\rm S}}b_{\rm zpl} -\frac{i}{1+\mathcal{C}_{\rm S}}\frac{\Omega_{\rm THz}}{\gamma_{\rm v}}.
\end{equation}

Considering $\Omega_{\rm THz} = 0$, we obtain the vibrational population under weak ZPL driving
\begin{equation}
    |\langle{\hat{b}\rangle}|^2 = \frac{\mathcal{C}_{\rm S}}{(1+\mathcal{C}_{\rm S})^2} \frac{\Omega_{\rm zpl}^2}{\gamma_{0}\gamma_{\rm v}}.
\end{equation}
Considering the values given in the main text of $\mathcal{C}_{\rm S} = 250$, $\gamma_{0}/(2\pi)=0.04\,$GHz, $\gamma_{\rm v}/(2\pi)=10\,$GHz, and $\Omega_{\rm zpl}/(2\pi)=1\,$GHz we have $|\langle{\hat{b}\rangle}|^2\approx 0.01 \ll 1$.

Under the same conditions, at zero ZPL input $b_{\rm zpl} = 0$, for small input vibrational excitation amplitudes $\Omega_{\rm THz}\ll \gamma_{\rm v}$ we can approximate the solution of Eq. \eqref{eq:bistab_b} as 
\begin{equation}
|b_{ss}| \approx 2n_{\rm S}\frac{\mathcal{C}_{\rm S}}{1+\mathcal{C}_{\rm S}}\frac{\Omega_{\rm THz}}{\gamma_{0}}  
\end{equation}
Plugging this solution into Eq. \eqref{eq:bistab_rhoee}, we obtain 
\begin{equation}
    \langle{ \hat{\sigma}^{\dag}\hat{\sigma} \rangle}_{ss} \approx  2n_{\rm S}\frac{\mathcal{C}_{\rm S}^2}{(1+\mathcal{C}_{\rm S})^2} \left(\frac{\Omega_{\rm THz}}{\gamma_{0}}\right)^2.
\end{equation}

\section{THz cross section enhancement in a resonant meta-material}
\label{app:thz_metamaterial}

As discussed in the main text, the only way to couple the molecule to an external THz radiative field is to collect the THz radiation with an antenna and then to compress it into a sub-wavelength region, where it can finally couple with the molecule.
This configuration could be naturally implemented by placing the molecules in a resonant meta-material. 
These structures are typically characterized by a gap of size $d\sim 1-10\,\mu$m where the oscillating electric field is confined \cite{maissen_ultrastrong_2014, liu_highly_2015, rajabali_ultrastrongly_2022}.
Even if we focus here on a specific regime, our theory is fully general and can be applied as well to IR transitions coupled to plasmonic structure, surface-enhanced Raman scattering phenomena, or other hybrid structures \cite{schmidt_quantum_2016,roelli_molecular_2016,martinez-garcia_coherent_2024, groiseau_single-photon_2024, jaber_hybrid_2024}.

\subsection{Capacitive coupling}
To model the combined system, we start by noticing that a THz cavity meta-material can be effectively modeled as a simple LC circuit with parallel plates capacitor \cite{porterfield_resonant_1994, zhou_saturation_2005, tretyakov_geometrical_2007}.
A basic Hamiltonian description is then provided by a simple LC resonant circuit
\begin{equation}
    H_{\rm LC} = \frac{\hat{Q}^2}{2C} + \frac{\hat{\Phi}^2}{2L},
\end{equation}
where the charge and magnetic flux dynamical variables follow the canonical commutation relations $[\hat{\Phi},\hat{Q}] = i\hbar$ as in superconducting circuits \cite{vool_introduction_2017}.

The meta-material gap in this description becomes the space between the parallel plates of the capacitor, where $d$ is then their relative distance.
The molecules enclosed directly inside the capacitor (as illustrated in Fig. \ref{fig:9}(b)) are excited by the oscillating electric field of the structure.
On the other side, the molecules can trigger an oscillation in the cavity electric field by inducing some charges on the metallic walls of the structure.
This mechanism realizes what is called capacitive coupling between the molecule and the metamaterial. 
It is implemented in the Hamiltonian description shifting the capacitor charge by the molecule's induced charges, obtaining \cite{de_bernardis_cavity_2018}
\begin{equation}
    H_{\rm LC}  \longmapsto \frac{\left( \hat{Q} - \hat{Q}_{\rm in} \right)^2}{2C} + \frac{\hat{\Phi}^2}{2L}.
\end{equation}
The induced charge is, in general, a complicated function of the molecule dipole transition moment and the geometry of the metamaterial. However, in the simplest case of infinite perfectly conducting walls, we have \cite{de_bernardis_cavity_2018}
\begin{equation}
    \hat{Q}_{\rm in} \approx \frac{e \xi_{\rm v}}{d} \left( \hat{b} +  {\rm h.c.}\right).
\end{equation}

Introducing the LC-circuit annihilation/creation operators $\hat{c} = \hat{Q}/\sqrt{2\hbar\omega_{\rm LC}C} - i \sqrt{\omega_{\rm LC}C/(2\hbar)}\hat{\Phi}$ and assuming that the RWA is valid, we have that
\begin{equation}\label{eq:ham_LC_Vibro_RWA}
    H_{\rm LC} \approx \hbar \omega_{\rm LC} \hat{c}^{\dag}\hat{c} + \hbar g_C \left( \hat{b}\, \hat{c}^{\dag} +  {\rm h.c.}\right).
\end{equation}
Here we have introduced the meta-material circuit resonant frequency $\omega_{\rm LC}=1/\sqrt{LC}$, and the molecule-circuit capacitive coupling
\begin{equation}
    \hbar g_C = \sqrt{\frac{e^2}{2C}\hbar \omega_{\rm LC}}\frac{\xi_{\rm v}}{d}.
\end{equation}
In the simplest configuration of parallel plates capacitor with area $A$, the capacitance is fully geometrical, given by $C=\epsilon_0 A/d$. In this case we can see that $\hbar g_C = \sqrt{e^2/(2\epsilon_0)\hbar\omega_{\rm LC}/V_{\rm LC}}\xi_{\rm v}$, with the usual dependence from the resonator volume $V_{\rm LC}=Ad$. 
It is worth stressing that describing the coupling in terms of the capacitance of the resonator $C$ is particularly convenient because can treat also the most generic case, when the resonator mode's volume could be difficult to access.

As a last step, we introduce a direct drive of the LC circuit by considering the term
\begin{equation}\label{eq:ham_LC_drive}
    H_{\rm LC-drive} =  \hbar\Omega_{\rm LC}\cos (\omega_{\rm THz} t) \left( e^{i\varphi}\hat{c} + e^{-i\varphi}\hat{c}^{\dag} \right).
\end{equation}
The LC-Rabi frequency $\Omega_{\rm LC}$ is determined by the absorption/reflection coefficient of the metamaterial and can be easily boosted by coupling it to a supplemental antenna \cite{jeannin_absorption_2020,jeannin_high_2020}. 
The phase $\varphi$ depends on whether the absorption occurs mostly due to the electric or magnetic field and, without loss of generality, we fix $\varphi = 0$.

\subsection{Metamaterial-enhanced vibrational driving}
\label{app:meta_gain}
We consider here the equations of motion deriving from coupling the molecule to the LC metamaterial through its vibrational dipole moment using the Hamiltonians in Eqs. \eqref{eq:ham_LC_Vibro_RWA}-\eqref{eq:ham_LC_drive} together with the Stokes Hamiltonian in Eq. \eqref{eq:ham_JC}
\begin{equation}
    \begin{split}
        i\partial_t \hat{b} & = \left(\Delta_{\rm v} - i\frac{\gamma_{\rm v}}{2}\right)\hat{b} + g_C\hat{c} + \frac{g_{\rm S}}{2}\hat{\sigma},
        \\
        i\partial_t \hat{c} & = \left(\Delta_{\rm LC} - i\frac{\gamma_{\rm LC}}{2}\right)\hat{c} + g_C\hat{b} + \frac{\Omega_{\rm LC}}{2}.
    \end{split}
\end{equation}
This equations are derived from the full master equation description including full dissipations and assuming the RWA in Eq. \eqref{eq:ham_LC_drive}, switching to a rotating frame where $\Delta_{\rm LC} = \omega_{\rm LC}-\omega_{\rm THz}$.
Here we have also included the dissipation for the LC circuit, for its dissipation rate is linked to its resistance $\gamma_{\rm LC} = R C$ (or $\gamma_{\rm LC} = R/L$ dependently if the decay channel is included in parallel or in series in the circuit scheme).
Since $\gamma_{\rm LC}/(2\pi)\sim O(1)\,$THz in typical subwavelength metamaterial structures at $\omega_{\rm LC}/(2\pi)\sim O(1)\,$THz (the quality factor is at most order 10) \cite{todorov_thz_2024}, we have that $\gamma_{\rm LC} \gg \Delta_{\rm v}, \Delta_{\rm LC}, g_C, g_{\rm S}$. Under such circumstances, we can adiabatically eliminate the metamaterial degree of freedom $\hat{c}$ by setting $\partial_t \hat{c}=0$.
We thus obtain 
\begin{equation}
    \hat{c} \approx - \frac{\Omega_{\rm LC}}{2(\Delta_{\rm LC} - i\gamma_{\rm LC}/2 )} - \frac{g_C}{\Delta_{\rm LC} - i\gamma_{\rm LC}/2  }\hat{b}.
\end{equation}
By inserting this expression in the other equation above, we have
\begin{equation}
    i\partial_t \hat{b} \approx \left[\Delta_{\rm v} + \delta\omega_C - \frac{i}{2}\left(\gamma_{\rm v} + \Gamma_C \right)\right]\hat{b} + \frac{g_{\rm S}}{2}\hat{\sigma} + \frac{\Omega_{\rm eff}}{2}.
\end{equation}
Here, we introduced the capacitive frequency shift, dissipation, and effective driving
\begin{equation}
    \begin{split}
        \delta\omega_{C} & = - \frac{g_C^2\Delta_{\rm LC}}{\Delta_{\rm LC}^2 + \gamma_{\rm LC}^2/4},
        \\
        \Gamma_C & = \frac{4g_C^2}{\gamma_{\rm LC}}\frac{\gamma_{\rm LC}^2/4}{\Delta_{\rm LC}^2 + \gamma_{\rm LC}^2/4},
        \\
        \Omega_{\rm eff} & = - g_C\Omega_{\rm LC} \frac{\Delta_{\rm LC}}{\Delta_{\rm LC}^2 + \gamma_{\rm LC}^2/4} + i g_C\Omega_{\rm LC} \frac{\gamma_{\rm LC}/4}{\Delta_{\rm LC}^2 + \gamma_{\rm LC}^2/4}.
    \end{split}
\end{equation}

Assuming the THz-LC resonance condition $\Delta_{\rm LC}=0$ we immediately have $\delta\omega_C = 0$ and $\Gamma_C = 4g_C^2/\gamma_{\rm LC}\approx 0$.
Using Eq. \eqref{eq:intensity_Efield} to redefine $\Omega_{\rm THz}$ we arrive to
\begin{equation}
    |\Omega_{\rm eff}| = 2g_C\frac{\Omega_{\rm LC}}{\gamma_{\rm LC}} = \Omega_{\rm THz} G_C,
\end{equation}
where we introduced the \emph{capacitive gain} factor
\begin{equation}
    G_C = \sqrt{\frac{\mathcal{Q}_{\rm LC}}{4\pi\alpha_{\rm fs}} \frac{\sigma_{\rm LC}}{d^2} \frac{2e^2}{C}\frac{1}{\hbar \omega_{\rm LC}} }.  
\end{equation}
Here, we have introduced the LC-metamaterial absorption cross section $\sigma_{\rm LC}$, and we have linked it to the LC Rabi frequency using Eqs. \eqref{eq:dipole_cross_section}-\eqref{eq:intensity_Efield}. Moreover we have introduced the LC-quality factor $\mathcal{Q}_{\rm LC}=\omega_{\rm LC}/\gamma_{\rm LC}$. 
From the previous equation, we can derive the molecule's effective THz drive after adiabatically eliminating the LC circuit
\begin{equation}
    \tilde{H}_{\rm THz} = G_C\hbar \Omega_{\rm THz}\left( \hat{b} + \hat{b}^{\dag} \right).
\end{equation}

\subsection{Capacitive gain experimental estimation}
Notice that by introducing the vacuum impedence $Z_{\rm vac} =1/(c\epsilon_0)\approx 377$ Ohm we can rewrite the capacitive gain in an even more compact format
\begin{equation}\label{eq:capacitive_gain}
    G_C = \sqrt{ 2 \mathcal{Q}_{\rm LC} \frac{\sigma_{\rm LC}}{d^2} \frac{Z_{\rm LC}}{Z_{\rm vac}} },
\end{equation}
where $Z_{\rm LC} = \sqrt{L/C} = 1/(C\omega_{\rm LC})$.
We remind here that $1\,{\rm Ohm}=1\,{\rm pF^{-1}THz^{-1}}$.

Interestingly, in the capacitive gain in Eq. \eqref{eq:capacitive_gain}, the capacitive gain of the cross section is not compared to the whole size of the structure but only to its capacitive gap size $d$ (or distance between the capacitor electrodes).
This is a direct consequence of the sub-wavelength compression of the electric field, leading to a notorious boost in the light-matter coupling strength in the THz range \cite{schlawin_cavity-mediated_2019,enkner_enhanced_2024, todorov_thz_2024}.

The dependence from the quality factor is instead reminiscent of the physics of the Purcell effect \cite{purcell_resonance_1946} but involving the absorption instead of the emission. For a typical THz structure we have $\mathcal{Q}_{\rm LC} \sim 10$ \cite{todorov_thz_2024}.

The last term in Eq. \eqref{eq:capacitive_gain} is instead the LC-impedance contribution. Assuming a capacitance $C\sim 320\,$aF \cite{tretyakov_geometrical_2007,jeannin_ultrastrong_2019, jeannin_quasi-static_2020}, we obtain an impedance $Z_{\rm LC}\approx 100\,{\rm Ohm}$ and an impedance ratio of $Z_{\rm LC}/Z_0 \approx 0.27$.

Collecting all these estimates all together and assuming a cross section $\sigma_{\rm LC} \sim 90\,\mu$m over a gap distance $d\sim 1\,\mu$m, we have that $G_C \approx 22$.
Even if this structure's cross section is quite large with respect to its current realizations \cite{jeannin_absorption_2020}, this value can in principle grow up to the maximum $\sigma_{\rm LC}^{\rm max} = 3/(8\pi)\lambda^2_{\rm LC}$ \cite{tretyakov_maximizing_2014}, where $\lambda_{\rm LC} = 2\pi c/\omega_{\rm LC}$. Having $\lambda_{\rm LC}\approx 60\,\mu$m we have a maximum value of $\sigma_{\rm LC}^{\rm max}\approx 430\,\mu$m$^2$.
Assuming an even smaller capacitance (so high impedance resonator) of $C\approx 86\,$aF, we have $Z_{\rm LC}\approx 375\,$Ohm and $Z_{\rm LC}/Z_0 \approx 1$.
With these more favorable parameters, we can reach the maximum capacitive gain at $G_C^{\rm max} \sim 100$.

To summarize, enclosing the molecule in a resonant THz metamaterial can boost the electric field driving the molecule by a factor $10^2$ thanks to the sub-wavelength field compression in the meta-material. 

Generalizing to $N_{\rm mol}$ molecules, the capacitive gain is boosted by the collective enhancement factor $\sqrt{N_{\rm mol}}$ due to the coupling to the same LC mode and the same Stokes laser. Together with an improvement of the design of the meta-material structure, increasing the cross sections and the impedance, this analysis shows that such a system can be a viable route toward the implementation of a THz single photon detector.

\bibliographystyle{mybibstyle}
 
\bibliography{references}

\end{document}